%% file: main.tex
\documentclass[sigconf,screen,nonacm]{acmart}

\usepackage{minted}
\newenvironment{longlisting}{\captionsetup{type=listing}}{}

\makeatletter
\let\orgdescriptionlabel\descriptionlabel
\renewcommand*{\descriptionlabel}[1]{%
  \let\orglabel\label
  \let\label\@gobble
  \phantomsection
  \edef\@currentlabel{#1}%
  \let\label\orglabel
  \orgdescriptionlabel{#1}%
}
\makeatother

\copyrightyear{2025}
\setcopyright{cc}
\setcctype{by}

\begin{document}

\title{Nagare Media Ingest: A System for Multimedia Ingest Workflows}

\author{Matthias Neugebauer}
\orcid{0000-0002-1363-0373}
\affiliation{%
  \institution{University of M\"unster}
  \city{M\"unster}
  \country{Germany}
}
\email{matthias.neugebauer@uni-muenster.de}

\begin{abstract}

  Ingesting multimedia data is usually the first step of multimedia workflows. For this purpose, various streaming protocols have been proposed for live and file-based content. For instance, SRT, RIST, DASH-IF Live Media Ingest Protocol and MOQT have been introduced in recent years. At the same time, the number of use cases has only proliferated by the move to cloud- and edge-computing environments. Multimedia systems now have to handle this complexity in order to stay relevant for today's workflows.

  This technical report discusses implementation details of \texttt{nagare media ingest}, an open source system for ingesting multimedia data into multimedia workflows. In contrast to existing solutions, \texttt{nagare media ingest} splits up the responsibilities of the ingest process. Users configure multiple concurrently running components that work together to implement a particular ingest workflow. As such, the design of \texttt{nagare media ingest} allows for great flexibility as components can be selected to fit the desired use case.

\end{abstract}



\maketitle

\input{body}

\bibliographystyle{ACM-Reference-Format}
\bibliography{references}

\end{document}

%% file: body.tex
\section{Introduction}
\label{sec:introduction}

Multimedia workflows have shifted from specialized appliances to software running on cloud- and edge-infrastructure. Data is now transferred over the network for which a variety of protocols have been proposed. According to the \textit{Annual Bitmovin Video Developer Report} from the year 2025---a survey among video developers---Real-Time Messaging Protocol~(RTMP)~\cite{hparmar_adobesrealtime_2012}, Secure Reliable Transport~(SRT)~\cite{sharabayko_srtprotocol_2021} and HTTP Live Streaming~(HLS)~\cite{pantos_httplivestreaming_2017} are the most used protocols for ingesting live content~\cite{bitmovininc_8thannualbitmovin_2025}. Still, new protocols are being introduced. Reliable Internet Stream Transport~(RIST)~\citep{videoservicesforum_tr0612020reliable_2020,videoservicesforum_tr0622021reliable_2021,videoservicesforum_tr063reliableinternet_2021} and Media over QUIC Transport~(MOQT)~\cite{nandakumar_mediaquictransport_2025} are two recent examples. And the DASH-IF Live Media Ingest Protocol~\cite{dashif_dashiflivemedia_2024} standardized how Dynamic Adaptive Streaming over HTTP~(DASH)~\cite{isoiec_isoiec230091_2022} and HLS, protocols typically deployed for distribution, can be used for ingest. Additionally, it standardized direct ingest of media in the Common Media Application Format~(CMAF)~\cite{isoiec_isoiec2300019_2024}. Multimedia systems for ingest workflows therefore potentially have to implement a variety of protocols. At the same time, there are different use cases ingest systems have to handle. Low-latency streaming, user-generated content, edge streaming, content protection, ad insertion or format transmuxing are just a few examples. If some functionality is missing, the whole system may be unsuitable for that particular use case. Ingest systems should therefore be designed with flexibility in mind. They should be able to evolve easily with the changing demands of users.

This technical report outlines the requirements, design decisions and implementation of the \texttt{nagare media ingest} research prototype~\cite{neugebauer_nagaremediaingest_2022}. We designed \texttt{nagare media ingest} to meet today's demands on ingest workflows. For that purpose, we split the ingest process into concurrently running components. In this way, we allow users to choose and configure components based on the required functionality. As an example, we prototypically implemented the DASH-IF ingest protocol as well as additional composable functionality.

The rest of this technical report is structured as follows. Section~\ref{sec:related-work} discusses related work. Next, Section~\ref{sec:overview-of-nagare-media-ingest} gives an overview of \texttt{nagare media ingest}. Sections~\ref{sec:volume}, \ref{sec:server}, \ref{sec:event-streaming}, \ref{sec:application} and~\ref{sec:function} go into more details about the volume, server, event, application and function components, respectively. Finally, Section~\ref{sec:conclusion} concludes this report.

\section{Related Work}
\label{sec:related-work}

\texttt{nagare media ingest} implements DASH-IF ingest as an example protocol including direct CMAF ingest. \textsc{\citeauthor{aguilararmijo_dynamicsegmentrepackaging_2020}} previously already proposed the use of CMAF for dynamic repackaging at the edge depending on the final delivery protocol. Because CMAF is maintained as the media format starting from the origin server over the content delivery network~(CDN) until the edge, they observed bandwidth savings, more cache hits and a reduced storage usage in an analytical model~\cite{aguilararmijo_dynamicsegmentrepackaging_2020}. This work focused on video-on-demand~(VoD) and streaming for final delivery. DASH-IF ingest, on the other hand, is primarily designed for live content.

\textsc{\citeauthor{mekuria_toolslivecmaf_2020}} present tools implementing the client-side of the DASH-IF ingest protocol in \citep{mekuria_toolslivecmaf_2020,unifiedstreaming_unifiedstreamingfmp4ingest_2022}. They also showcase a simple server receiving the stream and storing it as a CMAF track file. A~production-ready server implementation is only provided by the commercial Unified Origin product from Unified Streaming~\cite{jamiefletcher_livemediaingest_2020}. As far as we know, \texttt{nagare media ingest} is the only open source server implementation.

The DASH-IF ingest protocol has been adopted in other standards. The 3rd Generation Partnership Project~(3GPP) incorporated DASH-IF ingest into the 5G Media Streaming~(5GMS) specification for push-based content distribution~\cite{etsi_etsits126_2025}. As such, this protocol can play a vital role in multimedia streaming over 5G networks.

\section{Overview of \texttt{nagare media ingest}}
\label{sec:overview-of-nagare-media-ingest}

This section provides an overview of \texttt{nagare media ingest}. In Subsection~\ref{subsec:design}, we start with a bird's-eye view of the design with references to later sections for further details. Afterwards, we elaborate on our implementation in Subsection~\ref{subsec:implementation}. Lastly, Subsection~\ref{subsec:usage-and-configuration} discusses usage and configuration of \texttt{nagare media ingest}.

\subsection{Design}
\label{subsec:design}

When designing \texttt{nagare media ingest}, a system for multimedia ingest workflows in cloud and edge environments, we were guided by the following high-level requirements:

\begin{description}
  \item[R1\label{req:servers-apps}] An administrator should be able to configure various supported ingest protocols to run concurrently and side-by-side.
  \item[R2\label{req:functions}] An administrator should be able to configure additional supported functionalities for the configured ingest protocols to achieve the desired use case.
  \item[R3\label{req:functions-dev}] A developer should be able to implement additional functionality for the desired use case independently of the used ingest protocol.
  \item[R4\label{req:volume}] An administrator should be able to configure how and where ingested data is stored.
\end{description}

As mentioned in the introduction to this technical report, a variety of ingest protocols have been introduced. Users thus can choose the protocol based on the best fit for their use case. This requires a design that supports the implementation of multiple protocols. At the same time, we did not want to limit users to choose only one protocol at a time. Instead, multiple protocols should be able to run side-by-side (\ref{req:servers-apps}), thus supporting a variety of clients.

Ingest protocols usually only specify the transport of data. Depending on the use case, additional functionalities might be required to fully implement the ingest workflow. For instance, incoming media might need to be transmuxed into a different container format or encrypted for content protection. We expect that there are many of such additional functionalities necessary for various use cases. Consequently, we aim to have a design that makes it easy for developers to implement additional functionalities (\ref{req:functions-dev}) and for admins to configure them (\ref{req:functions}). At the same time, note that \texttt{nagare media ingest} is not a full workflow system. Additional functionality should only operate in the immediate context of an ingest. \texttt{nagare media ingest} should facilitate passing on more complex tasks to dedicated workflow systems.

The storage requirements also vary. Data might only need to be stored in memory for a short time. Other use cases might store it on the local filesystem or in a cloud storage such as S3\footnote{An object storage system introduced by Amazon Web Services (see \url{https://aws.amazon.com/s3/}) and reimplemented by other cloud providers and software systems.}. The design should enable different storage implementations (\ref{req:volume}).

\begin{figure}[h]
  \centering
  \includegraphics[width=0.8\columnwidth]{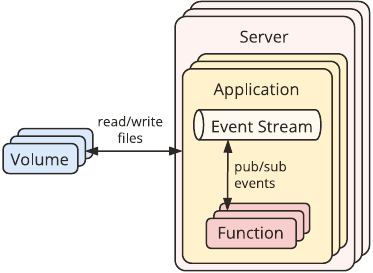}
  \caption{Ingest server design, compare~\cite{neugebauer_nagaremediaingest_2022}.}
  \label{fig:nmi-design}
  \Description{Ingest server design.}
\end{figure}

To meet these requirements, we propose a design as depicted in Figure~\ref{fig:nmi-design}. Here, we split the implementation into different concurrently running components that are responsible for specific areas of the ingest process. This design allows evolving the implementation more easily with future changes in protocols and use cases as only affected components need to be adapted.

First, we have the server component that is listening for and establishing connections to clients. Multiple servers can run side-by-side each listening on a different address (i.e.~network interfaces and/or port number). The server component allows extracting the logic of ingest protocols that share a common transport protocol. For instance, protocols based on HTTP could rely on an HTTP server.

Next, each server is running one or more applications. This component implements the basic logic of the ingest protocol, i.e.~accepting the request from the client followed by transferring and storing the incoming data. Depending on the capabilities of the protocol, client requests could be routed to applications based on different factors. For instance, an HTTP server could match applications by domain names (virtual host) or paths. This design thus additionally enables multi-tenancy scenarios. Note that we used the more generic term application to emphasize that non-ingest applications, e.g.~to retrieve ingested data or expose metrics, are also a possibility.

Part of every application is an in-memory event stream. Applications should emit relevant events, for instance, when an ingest starts or stops. Other components can subscribe to this event stream and react to relevant events. This indirection results in a loose coupling between components and a synchronization in the multithreaded design. Events are typed and can contain further properties and references to other objects.

With the volume component, other components have an abstract way to store and retrieve ingested data. The interface is file-oriented, i.e.~components can open files for writing or reading as well as deleting files. However, volume implementations can implement this interface for various underlying storage locations. Volumes should provide certain guarantees as defined in the interface in order to have a predictable outcome (see Section~\ref{sec:volume}).

Lastly, any additional functionally should be implemented in function components. They are associated with one specific application and run concurrently. Functions typically subscribe to the application's event stream and react to certain events. However, they may also emit events themselves allowing multiple functions to work towards a specific use case. Moreover, functions can make use of volumes to store and retrieve files.

\begin{figure}[h]
  \centering
  \includegraphics[width=0.8\columnwidth]{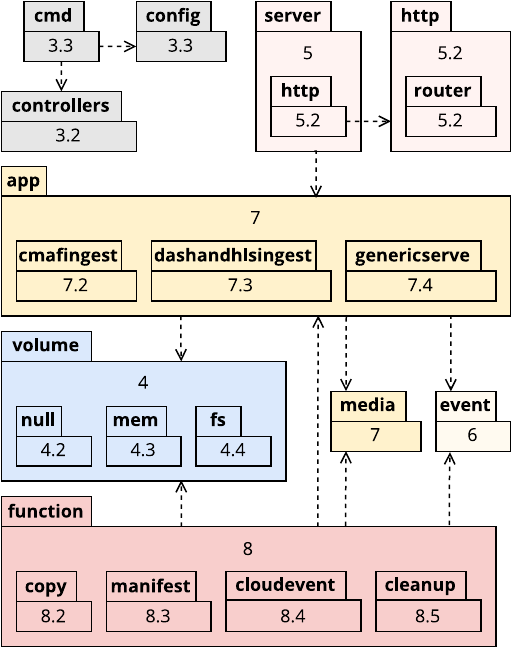}
  \caption{Overview of important packages and interdependencies.}
  \label{fig:nmi-uml-packages}
  \Description{Overview of important packages and interdependencies.}
\end{figure}

The design of \texttt{nagare media ingest} is reflected in its software packages. Figure~\ref{fig:nmi-uml-packages} depicts the most important packages and their interdependencies. More details are provided in the coming sections as referred to by the numbers within the packages. Note that the \texttt{config} package bundles types for configuring components and hence many packages depend on it. For increased readability, we refrained from depicting the dependencies to this package in this report. Moreover, we omitted smaller, less important packages. We applied our design and provided example implementations in our research prototype (see depicted subpackages). In particular, we partially implemented the DASH-IF ingest protocol as will be discussed in subsequent sections.

\subsection{Implementation}
\label{subsec:implementation}

We implemented our design in the open source software \texttt{nagare media ingest} and published it to \url{https://github.com/nagare-media/ingest} under the Apache~2.0 license. We used Go as programming language for several reasons. First, our design is highly multithreaded. Go provides robust support for concurrency with Goroutines, a green thread model approach. Moreover, Go's channel construct provides a simple mechanism to pass data between threads. Go supports many hardware architectures and operating systems. Finally, Go is a widespread language in cloud and edge environments letting \texttt{nagare media ingest} fit into the larger ecosystem.

Our repository includes a \texttt{Makefile} to help develop and build the project. \texttt{nagare media ingest} can be cross-compiled to various operating systems and hardware architectures. It is known to run well on AMD64 and ARM64 within Linux and macOS. Additionally, the \texttt{Makefile} contains targets to build production-ready Linux container images that are easily deployable to container runtimes. Moreover, new container images are built automatically in a continuous integration pipeline and pushed to the GitHub container registry\footnote{\url{https://github.com/nagare-media/ingest/pkgs/container/ingest}}.

For our implementation, we relied on a number of software libraries. The most important ones are the following. As a server application, administrators configure \texttt{nagare media ingest} via command-line arguments as well as a configuration file. For parsing both, we used \texttt{github.com/spf13/pflag}\footnote{\url{https://github.com/spf13/pflag}} and \texttt{github.com/spf13\linebreak{}/viper}\footnote{\url{https://github.com/spf13/viper}}, respectively. Additionally, \texttt{github.com/go-viper/map\linebreak{}structure/v2}\footnote{\url{https://github.com/go-viper/mapstructure}} allows easily mapping the parsed configuration to Go structures. Structured logging is handled by \texttt{go.uber.org/zap}\footnote{\url{https://github.com/uber-go/zap}} to inform administrators about the current state. For our HTTP server, we used \texttt{github.com/gofiber/fiber/v2}\footnote{\url{https://github.com/gofiber/fiber/tree/v2}} as the web framework that builds upon \texttt{github.com/valyala/fasthttp}\footnote{\url{https://github.com/valyala/fasthttp}} for the HTTP server implementation. Lastly, \texttt{github.com/Eyevinn/\linebreak{}mp4ff}\footnote{\url{https://github.com/Eyevinn/mp4ff}} is used for parsing media in the ISO Base Media File Format~(ISO~BMFF)~\cite{isoiec_isoiec1449612_2022}.

\begin{figure*}[t]
  \centering
  \includegraphics[width=\textwidth]{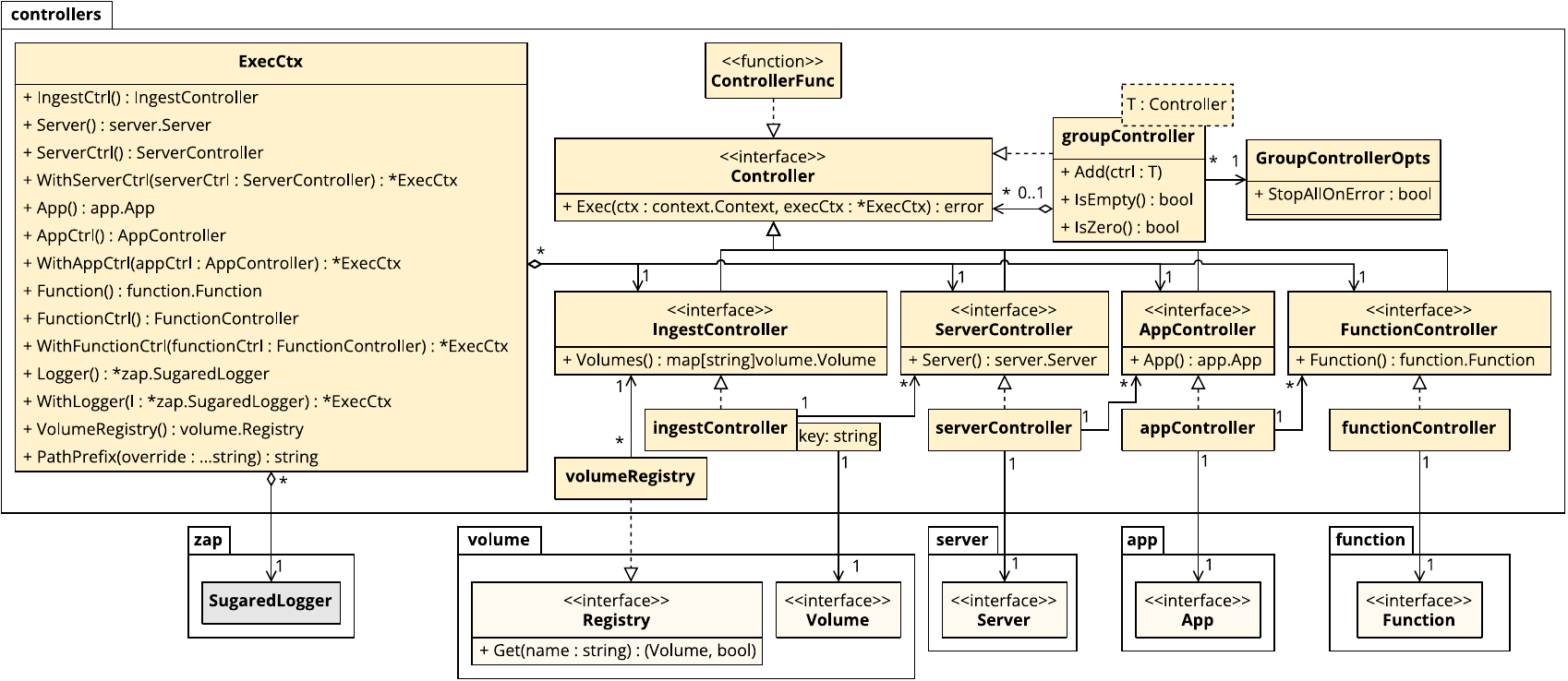}
  \caption{UML class diagram of the types in the \texttt{controllers} package.}
  \label{fig:nmi-uml-class-controller}
  \Description{UML class diagram of the types in the \texttt{controllers} package.}
\end{figure*}

Except for volumes, which are mostly passive, every other component has its own main thread from where potentially more threads can be spawned. To manage the initialization and execution of the components, a corresponding controller struct was implemented. The initialization and management of the whole system is implemented in an additional controller. Figure~\ref{fig:nmi-uml-class-controller} depicts the types in the \texttt{controllers} package.

The \texttt{Controller} interface only specifies the \texttt{Exec} method for starting the execution of the controller and the associated component. Here, we use a common Go pattern by passing a \texttt{Context} as the first argument. In Go, a \texttt{Context} provides a mechanism to pass down cancellation signals as well as context-dependent values. By consistently passing the \texttt{Context} throughout the code, we can tie the cancellation of the \texttt{Context} to the cancellation of the currently executed threads. Our \texttt{Controller} thus does not need a method for stopping the execution. As the second parameter, the \texttt{Exec} method takes a pointer to \texttt{ExecCtx}. This structure stores and provides access to relevant objects and is passed in the code similar to the Go \texttt{Context}. Moreover, there are derived controller interfaces for the relevant components: \texttt{ServerController}, \texttt{AppController}, \texttt{FunctionController} and the additional \texttt{IngestController} that initializes volumes and manages the system itself. Each interface specifies a method for returning the component under management. We implement each interface as a Go structure. Next to managing the associated component, it is also the responsibility of one controller to instantiate cascading controllers. For example, the \texttt{serverController} will create \texttt{appController} instances as configured by the administrator. We implemented a \texttt{groupController} structure to ease the handling of multiple controllers. Moreover, function values following a specific signature can be type converted to a \texttt{ControllerFunc} that also implements the \texttt{Controller} interface. Each component is named by the administrator. This is especially important when referencing volumes. The \texttt{volumeRegistry} structure provides a way for components to access volumes based on a given name.

\subsection{Usage and Configuration}
\label{subsec:usage-and-configuration}

\texttt{nagare media ingest} is a server application. As such, its operation is driven by the configuration passed to the process. First, some options are controlled by command-line arguments. However, we intentionally limited the configuration through command-line arguments to a few basic options. Instead, a configuration file mainly influences which components are instantiated and how they are configured. Listing~\ref{lst:configuration} demonstrates how a configuration file might look like.
\begin{longlisting}
\begin{minted}{yaml}
apiVersion: ingest.nagare.media/v1alpha1
kind: Config

servers:
  - name: http
    address: :8080
    http:
      idleTimeout: 60s

    apps:
      - name: cmaf
        http:
          host: ingest.nagare.media
          path: /cmaf
        cmafIngest:
          volumeRef:
            name: fsVol
          streamTimeout: 5m

        functions:
          - name: manifest
            manifest:
              volumeRef:
                name: memVol

      - name: serve
        http:
          host: "**"
          path: /streams
          cors:
            allowOrigins: "*"
            allowMethods: GET,HEAD,OPTIONS
            allowHeaders: "*"
            maxAge: 86400
        genericServe:
          appRef:
            name: cmaf
          volumeRefs:
            - name: memVol
            - name: fsVol

volumes:
  - name: memVol
    mem: {}
  - name: fsVol
    fs:
      path: /tmp/nagare/ingest/volumes/fsVol
\end{minted}
  \caption{Example configuration file.}
  \label{lst:configuration}
  \Description{Example configuration file.}
  \vspace{10pt}
\end{longlisting}

Configuration files are written in the YAML data exchange language~\cite{benkiki_yamlaintmarkup_2004}. For the fields, we followed the approach of Kubernetes\footnote{\url{https://kubernetes.io}}, a container orchestration system. All configuration files therefore have an \texttt{apiVersion} and \texttt{kind} field in order to differentiate them from other YAML files. Additionally, versioned configuration files provide a safer way to introduce breaking changes in new releases.

\begin{figure*}[t]
  \centering
  \includegraphics[width=\textwidth]{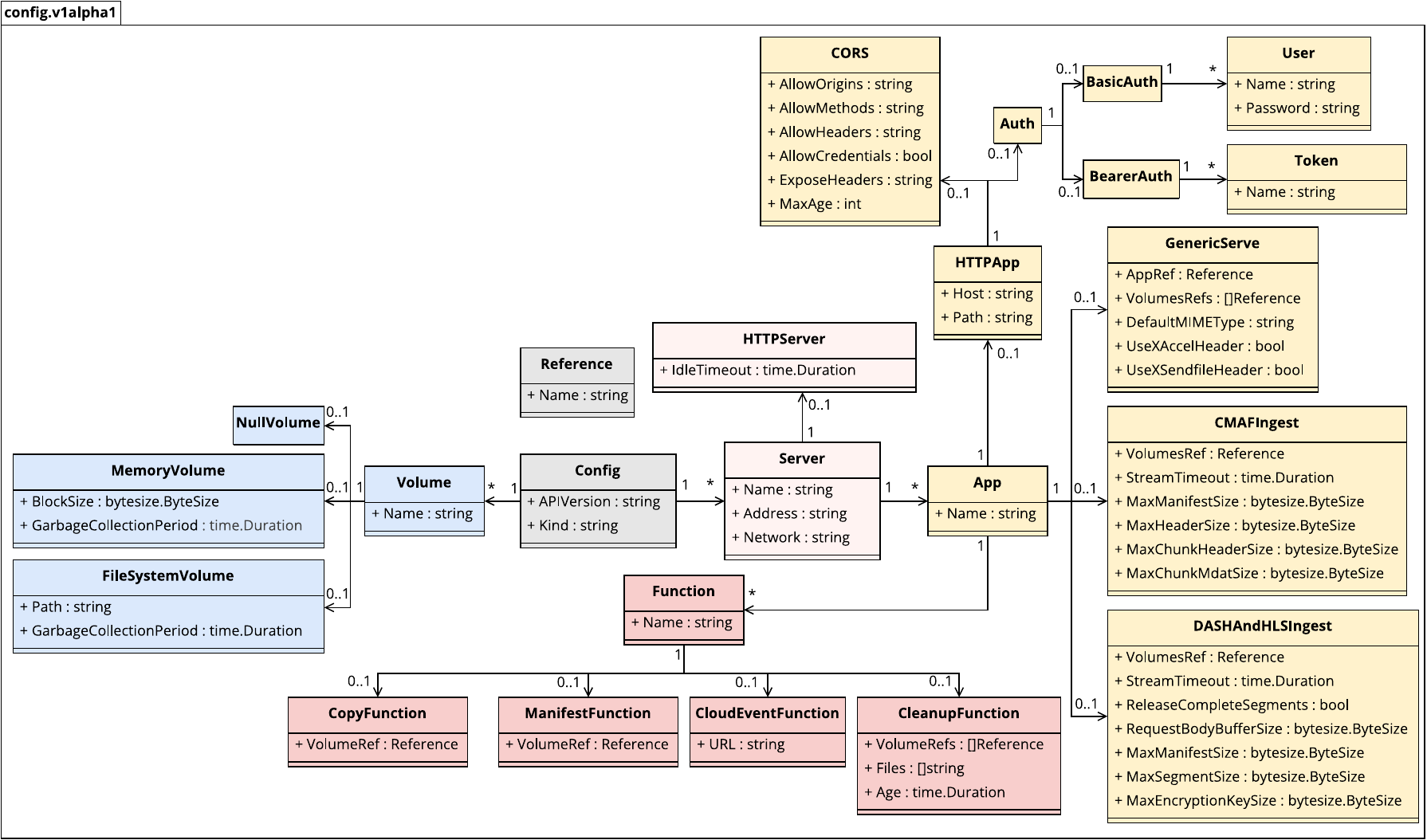}
  \caption{UML class diagram of the types in the \texttt{config.v1alpha1} package.}
  \label{fig:nmi-uml-class-config}
  \Description{UML class diagram of the types in the \texttt{config.v1alpha1} package.}
\end{figure*}

In this example, we configure one HTTP server (see Subsection~\ref{subsec:http}) named \texttt{http} that listens for new requests on all network interfaces on port 8080. We pass additional configuration to the server by setting the idle timeout. Next, we configure the two applications \texttt{cmaf} and \texttt{serve}. The first instantiates the \texttt{cmafIngest} application (see Subsection~\ref{subsec:cmafingest-dash-if-ingest-interface-1}) while the second instantiates \texttt{genericServe} (see Subsection~\ref{subsec:genericserve}). Requests coming to the host \texttt{ingest.nagare.\linebreak{}media} under the path \texttt{/cmaf} are mapped to the \texttt{cmaf} application. The \texttt{serve} application receives requests from any host under the \texttt{/streams} path. The applications reference the two volumes \texttt{memVol} and \texttt{fsVol} that are instances of \texttt{mem} (see Subsection~\ref{subsec:mem}) and \texttt{fs} (see Subsection~\ref{subsec:fs}), respectively. The \texttt{cmaf} application has an associated \texttt{manifest} function instance of the correspondingly named function (see Subsection~\ref{subsec:manifest}).

Most components provide configuration options. After parsing the YAML data, it is mapped to Go structures based on struct tags. Listing~\ref{lst:go-struct-tags} shows the type definition of the \texttt{Config} structure with struct tags (on the rightmost side of each field).
\begin{listing}[h]
  \begin{minted}[xleftmargin=20pt,linenos,fontsize=\small]{go}
type Config struct {
  APIVersion string   `mapstructure:"apiVersion"`
  Kind       string   `mapstructure:"kind"`
  Servers    []Server `mapstructure:"servers,omitempty"`
  Volumes    []Volume `mapstructure:"volumes,omitempty"`
}
  \end{minted}
  \caption{Example Go struct with struct tags.}
  \label{lst:go-struct-tags}
  \Description{Example Go struct with struct tags.}
  \vspace{10pt}
\end{listing}

In Go, struct tags are often used to configure serialization and deserialization. In this example, we specify the names of the fields as found in the YAML data. Additionally, the \texttt{omitempty} option is employed to omit ``empty'' fields when serializing (e.g.~lists without any elements). We define struct tags for each struct type in the \texttt{config.v1alpha1} package. Figure~\ref{fig:nmi-uml-class-config} depicts a class diagram of the structures used for mapping the parsed YAML data. These structures are later also passed as arguments when instantiating components.

\section{Volume}
\label{sec:volume}

This section discusses the volume component type. We start with an overview of the general interface in Subsection~\ref{subsec:volume-overview}. Then Subsections~\ref{subsec:null},~\ref{subsec:mem} and~\ref{subsec:fs} will detail the specific volume implementations \texttt{null}, \texttt{mem} and \texttt{fs}, respectively.

\subsection{Overview}
\label{subsec:volume-overview}

\begin{figure*}[t]
  \centering
  \includegraphics[width=0.8\textwidth]{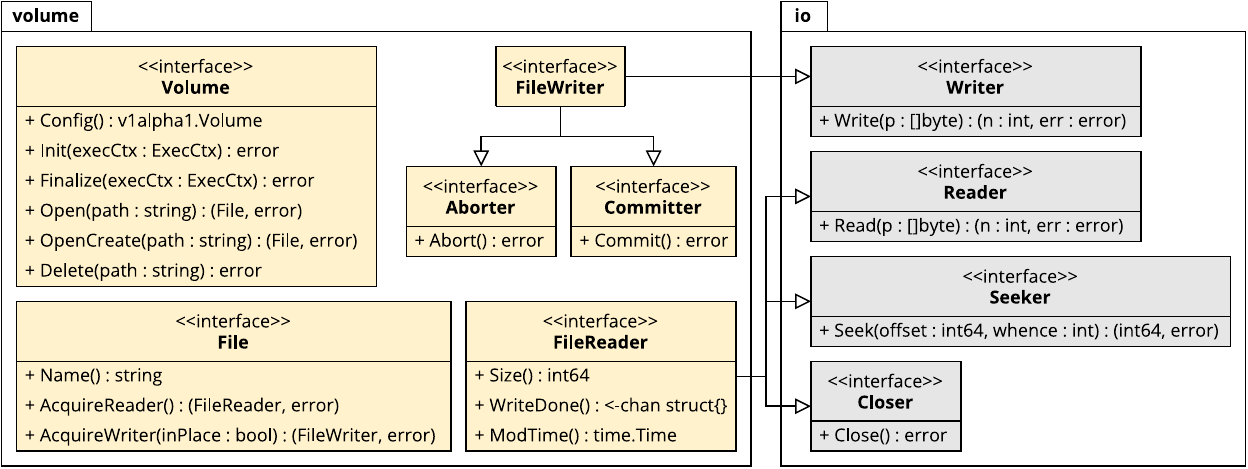}
  \caption{UML class diagram of the types in the \texttt{volume} package.}
  \label{fig:nmi-uml-class-volume}
  \Description{UML class diagram of the types in the \texttt{volume} package.}
\end{figure*}

The volume component provides a way for other components to store data. We adopt a file-based approach for our interfaces. Additionally, we extend generic I/O interfaces in the standard library to ensure compatibility with the Go ecosystem. Figure~\ref{fig:nmi-uml-class-volume} depicts the interface types in the \texttt{volume} package.

\texttt{Volume} specifies the interface of the volume component itself. Before using a volume, it might need an initialization. Similarly, when terminating \texttt{nagare media ingest}, volumes might need to run some cleanup logic. For instance, the storage containers in the underlying system might need to be created and later deleted. This logic can be implemented in the \texttt{Init} and \texttt{Finalize} methods and the \texttt{IngestController} (see Subsection~\ref{subsec:implementation}) will make sure to call them. The \texttt{Open} and \texttt{OpenCreate} methods both open the file under the given path, i.e.~return an implementation of \texttt{File}. While the first method will return an error if the file does not already exist, the second method will create a new file in such a case. Finally, the \texttt{Delete} method allows deleting files.

The \texttt{File} interface allows acquiring new readers and writers. While there can be multiple concurrent readers, implementations must make sure that at any time there is at most one writer. Furthermore, readers and a writer can run concurrently in one of two ways. If the acquired writer operates in-place, i.e.~the \texttt{inPlace} parameter is \texttt{true}, written data is immediately available to existing readers even before closing the writer. Additionally, readers will block and wait for the writer when they reach the end of the currently written data. This behavior is especially important in low-latency scenarios where immediate access to written data is crucial. Alternatively, writers can operate not in-place. Here, readers will return the data that was previously written to the file by the last finished writer at the time of acquiring the reader. If there was no previous writer, readers will read an empty file, i.e.~immediately return an end of file~(EOF) signal. The advantage of this approach is that readers will not be blocked by a writer at the cost of potentially stale data.

The \texttt{FileWriter} and \texttt{FileReader} interfaces represent writers and readers, respectively. \texttt{FileWriter} extends Go's generic \texttt{Writer} interface for writing to data. Additionally, it extends our \texttt{Aborter} and \texttt{Committer} interfaces. Both are used to close the \texttt{FileWriter}. While a call to \texttt{Commit} will adopt all written data to the file and make it available to readers, \texttt{Abort} will discard any written data.

\texttt{FileReader} extends the Go interfaces \texttt{Reader}, \texttt{Seeker} and \linebreak{}\texttt{Closer} for the corresponding actions. Moreover, the \texttt{Size} and \texttt{ModTime} attributes provide file metadata about the size and time of the last modification. Finally, the \texttt{WriteDone} method returns a Go channel that is closed when there is no writer anymore. Using a channel in this way is a common idiom in Go programs for checking the existence of other routines and signaling state changes.

\subsection{\texttt{null}}
\label{subsec:null}

\begin{figure}[h]
  \centering
  \includegraphics[width=\columnwidth]{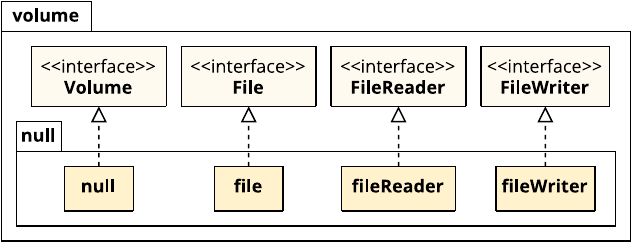}
  \caption{UML class diagram of the types in the \texttt{volume.null} package.}
  \label{fig:nmi-uml-class-volume-null}
  \Description{UML class diagram of the types in the \texttt{volume.null} package.}
\end{figure}

\texttt{null} is a dummy implementation of the volume interfaces. It will discard any written data and always return an EOF signal when reading files. We primarily provide this implementation for testing purposes. Figure~\ref{fig:nmi-uml-class-volume-null} depicts the relevant types.

As a dummy implementation, \texttt{null} is rather simple. All required interfaces are implemented and no complex coordination logic between readers and a writer is necessary.

\subsection{\texttt{mem}}
\label{subsec:mem}

\begin{figure}[h]
  \centering
  \includegraphics[width=\columnwidth]{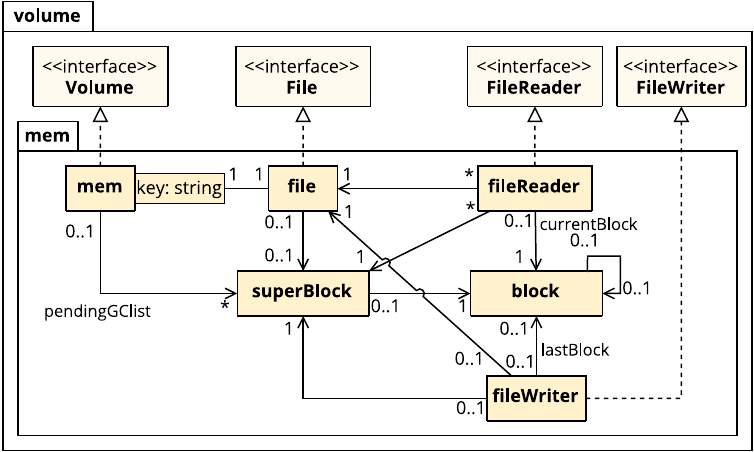}
  \caption{UML class diagram of the types in the \texttt{volume.mem} package.}
  \label{fig:nmi-uml-class-volume-mem}
  \Description{UML class diagram of the types in the \texttt{volume.mem} package.}
\end{figure}

\texttt{mem} is an in-memory implementation of the volume interfaces. This is especially advantageous in cases where there are files that are frequently accessed. HLS and DASH manifest files are examples in multimedia streaming scenarios. These small and constantly overwritten files can be kept in memory for quick access. In Figure~\ref{fig:nmi-uml-class-volume-mem}, the types in the \texttt{volume.mem} package are depicted.

We keep track of files in a \texttt{string} (the path) to \texttt{file} map. Creating a new file thus will add a new entry in this map. The file's content is stored as a linked list of \texttt{block} structures. Each \texttt{block} contains a pointer to the next \texttt{block} as well as a \texttt{byte} array. The size of a \texttt{block} can be configured by administrators and remains constant for the same \texttt{mem} instance. Choosing the right \texttt{block} size for efficient memory usage depends on the use case. For instance, MPEG transport stream~(MPEG-TS)~\cite{isoiec_recituth2220_2023}, a media container format, has a fixed 188-byte packet size. Typically, multiple packets are transferred and stored in concatenated form. A multiple of 188 would thus be ideal to avoid only partially filled \texttt{block}s. We recognize that this fine-tuning requires detailed knowledge on the side of the administrator. However, the use of linked equally sized \texttt{block}s allows creating files where the size is unknown beforehand. Moreover, we use memory pooling for retrieving \texttt{block}s, i.e.~reclaimed \texttt{block}s are handed back to the pool and can be reused first before allocating memory for new ones.

\begin{figure}[h]
  \centering
  \includegraphics[width=\columnwidth]{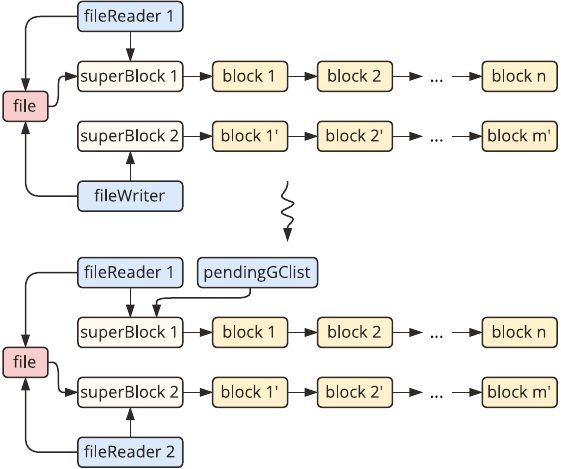}
  \caption{Illustration of committing not in-place written data to a file.}
  \label{fig:nmi-volume-mem-commit}
  \Description{Illustration of committing not in-place written data to a file.}
\end{figure}

The \texttt{superBlock} provides an entry to the linked list and additionally stores internal and metadata fields. A \texttt{file} then points to a \texttt{superBlock}. This indirection further allows for committing and aborting writers. Figure~\ref{fig:nmi-volume-mem-commit} illustrates this in an example scenario.

Here, \texttt{fileWriter} is not operating in-place: the \texttt{file} still points to the old \texttt{superBlock}~\texttt{1} while the \texttt{fileWriter} writes data to the \texttt{superBlock}~\texttt{2} list. Aborting the written data would just close the \texttt{fileWriter} for further writes and return the \texttt{block}s of \linebreak{}\texttt{superBlock}~\texttt{2} back to the pool. Committing the data, on the other hand, would result in the scenario depicted at the bottom of Figure~\ref{fig:nmi-volume-mem-commit}. The \texttt{file} now points to the new \texttt{superBlock}~\texttt{2}. Old readers such as \texttt{fileReader}~\texttt{1} can still continue reading. However, \texttt{superBlock}~\texttt{1} was added to the pending garbage collection list (\texttt{pendingGClist}) and will be reclaimed once all readers have been closed. At the same time, new readers such as \texttt{fileReader}~\texttt{2} already read from the new \texttt{superBlock}~\texttt{2}.

In our implementation, read-write mutexes are used to synchronize \texttt{fileReader}s and a \texttt{fileWriter}. Additionally, in-place writers can signal the availability of new data to all readers through Go channels. For the garbage collection of \texttt{superBlock}s, we keep track of open readers with an atomic counter.

\subsection{\texttt{fs}}
\label{subsec:fs}

\begin{figure}[h]
  \centering
  \includegraphics[width=\columnwidth]{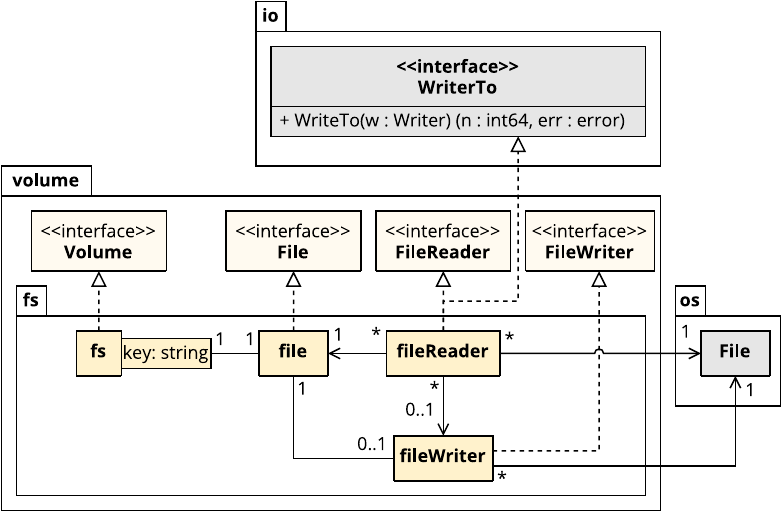}
  \caption{UML class diagram of the types in the \texttt{volume.fs} package.}
  \label{fig:nmi-uml-class-volume-fs}
  \Description{UML class diagram of the types in the \texttt{volume.fs} package.}
\end{figure}

With \texttt{fs}, we provide a volume implementation that stores files on the local filesystem. Administrators configure \texttt{fs} instances with a path to a directory. Files created in the volume will then be placed relative to this path. Figure~\ref{fig:nmi-uml-class-volume-fs} shows all relevant types in the \linebreak{}\texttt{volume.fs} package.

In order to coordinate writers and readers, \texttt{fs} needs to keep track of all open \texttt{file}s. Similar to \texttt{mem} (see Subsection~\ref{subsec:mem}), we use a string to \texttt{file} map for that. However, unlike \texttt{mem} where files obviously need to remain in this map until deleted, \texttt{fs} will remove entries regularly when a \texttt{file} is no longer open.

Each \texttt{fileReader} and \texttt{fileWriter} has its own file descriptor as represented by the associations with Go's \texttt{os.File} structure. Reads under an in-place writer will block when they reach the end of the currently written data. For this, readers can check how many bytes the writer has already written. We again use mutexes for synchronization and Go channels to signal the availability of new data and to unblock readers.

In addition to the \texttt{FileReader} interface, \texttt{fileReader} also implements Go's \texttt{WriterTo} interface. This enables an optimization for a typical scenario in \texttt{nagare media ingest}. Go has special handling on UNIX operating systems when the content of a file should be written to a socket. In this case, the \texttt{sendfile} system call is used to directly copy the data within the operating system kernel instead of passing data between kernel and user space. This behavior is possible thanks to a careful implementation of \texttt{WriterTo}. Additionally, we only trigger this optimization in the absence of an in-place writer as we need to handle coordination in this case. Still, sending complete files over the network, e.g.~for further processing after the ingest, is a typical scenario making this optimization worthwhile.

\texttt{fileWriter} implements aborting and committing written data by writing to another file which is then moved atomically. In-place writers first rename the existing file with a temporary name and create a new file in its place. Aborting the writer will move the old file back and thus restore the previous content. A commit, on the other hand, will delete the old file. Writers operating not in-place behave similarly, however, the new data is first written to a temporary file that is atomically moved over the existing one when committing or deleted when aborting.

\section{Server}
\label{sec:server}

This section discusses the server component. In Subsection~\ref{subsec:server-overview}, we first give an overview of the generic design and then detail our implementation for HTTP in Subsection~\ref{subsec:http}.

\subsection{Overview}
\label{subsec:server-overview}

\begin{figure}[h]
  \centering
  \includegraphics[width=0.7\columnwidth]{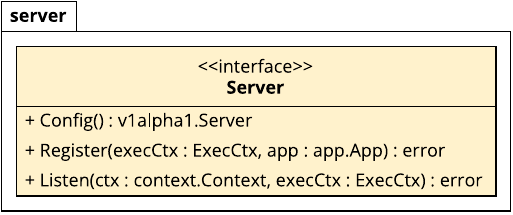}
  \caption{UML class diagram of the types in the \texttt{server} package.}
  \label{fig:nmi-uml-class-server}
  \Description{UML class diagram of the types in the \texttt{server} package.}
\end{figure}

The server component implements the basis for running ingest applications (see Section~\ref{sec:application}). Its main purpose is to open network ports and listen for requests. Accepted requests should then be delegated to the appropriate application. Figure~\ref{fig:nmi-uml-class-server} depicts the \texttt{Server} interface.

Servers must implement three methods. \texttt{Config} is just a getter for the server configuration at use. More important is \texttt{Register}, the method for deploying an application to this server. Currently, there is no way to remove a previously registered application. \texttt{nagare media ingest} only supports a static component configuration loaded during startup. Dynamically adding and removing applications could be implemented in the future. Servers have to make sure that only compatible applications are registered, e.g.~by checking if applications implement additional interfaces. \texttt{Listen}, the third method, starts the server. As detailed before, server termination is handled by the passed Go \texttt{Context} (see Subsection~\ref{subsec:implementation}).

\subsection{\texttt{http}}
\label{subsec:http}

\begin{figure*}[t]
  \centering
  \includegraphics[width=0.9\textwidth]{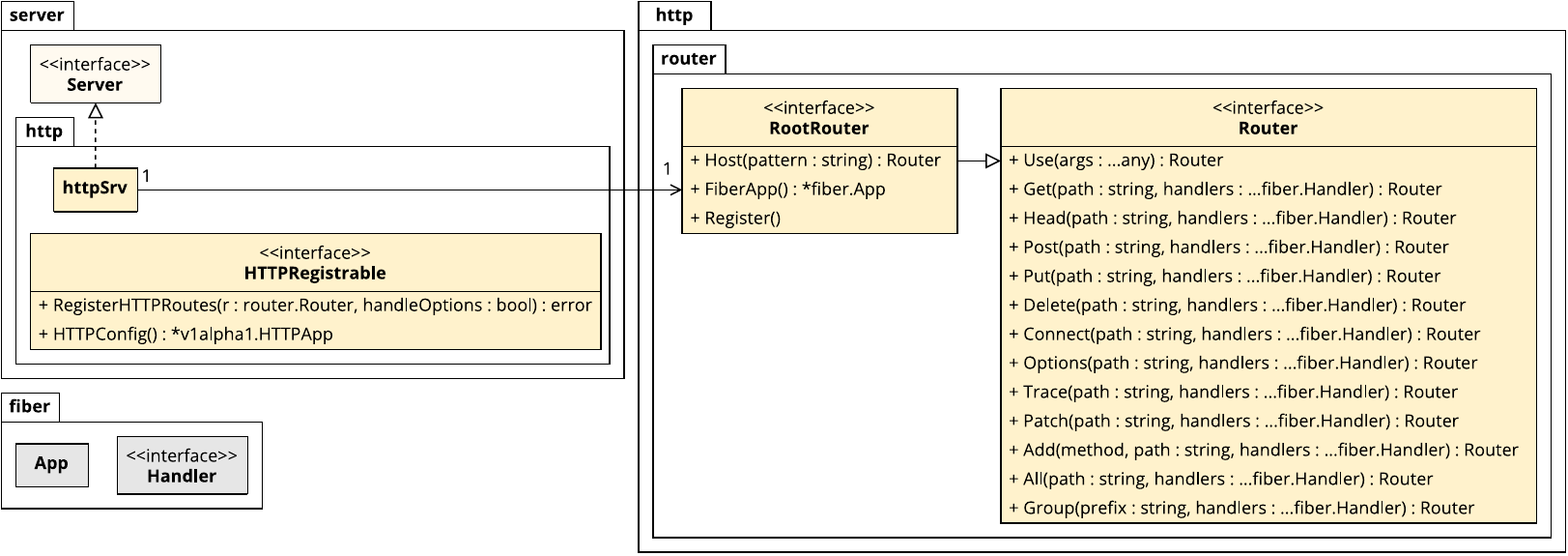}
  \caption{UML class diagram of relevant packages and types for the \texttt{http} server component.}
  \label{fig:nmi-uml-class-server-http}
  \Description{UML class diagram of relevant packages and types for the \texttt{http} server component.}
\end{figure*}

\texttt{nagare media ingest} includes an \texttt{http} server component capable of running HTTP/1.1-based ingest protocols. As of this report, this is the only available server component. We use \texttt{github.com/valyala\linebreak{}/fasthttp} (fasthttp) as HTTP/1.1 implementation in connection with \texttt{github.com/gofiber/fiber/v2} (Fiber) as lightweight web framework. Figure~\ref{fig:nmi-uml-class-server-http} shows relevant packages and types for the \texttt{http} server component.

Applications that can be registered to \texttt{httpSrv}, need to implement the \texttt{HTTPRegistrable} interface. Trying to register applications that do not implement that interface will fail. It prescribes two methods. With \texttt{HTTPConfig}, applications can provide general configuration used by \texttt{httpSrv} for handling requests to this application. Most importantly, it sets the mounting path of the application, i.e.~a path prefix for all requests to this application. Moreover, a host pattern can be configured for host-based request routing. For this, we use the library \texttt{github.com/gobwas/glob}\footnote{\url{https://github.com/gobwas/glob}} for UNIX-like globbing to match a pattern to the \texttt{Host} request header. For instance, the pattern \texttt{"*.example.com"} would match \texttt{primary.example.com} as well as \texttt{backup.example.com} but not \texttt{primary.example.org}. The wildcard \texttt{*} matches for a single word, i.e.~any string between dots. To match for any number of words, the \texttt{**} wildcard can be used. Applications with the host pattern \texttt{"**"} would therefore receive requests from any host. For single characters, the \texttt{?} wildcard exists. We additionally support character lists and ranges as well as pattern alternatives. Next to the mounting path and host pattern, the configuration also includes options for request authentication (e.g.~HTTP Basic Authentication~\cite{reschke_basichttpauthentication_2015}) as well as Cross-Origin Resource Sharing~(CORS)~\cite{whatwg_fetch_2025}.

\begin{figure}[H]
  \centering
  \includegraphics[width=0.7\columnwidth]{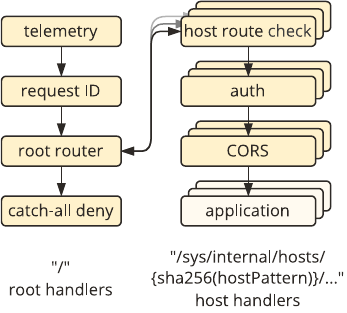}
  \caption{Request flow from root to host handlers and finally to the application in the \texttt{http} server component.}
  \label{fig:nmi-http-request-handling}
  \Description{Request flow from root to host handlers and finally to the application in the \texttt{http} server component.}
\end{figure}

The second method, \texttt{RegisterHTTPRoutes}, is called by \texttt{httpSrv} to pass a \texttt{Router} implementation to the application. The \texttt{Router} is then used by the application to map HTTP verbs and paths to request handlers. As such, the \texttt{Router} has methods corresponding to HTTP verbs such as \texttt{GET} or \texttt{POST}. It adapts and extends Fiber's built-in path-based router, e.g.~in order to implement the host-based request routing described above. For this, we remap routes defined by applications to internal paths and implement internal redirects with custom matching logic. In Fiber, request handlers are chained and multiple handlers can be mapped to the same path. Each handler can work on the request and then respond or pass the request to the next handler. This effectively allows implementing HTTP middlewares. In \texttt{nagare media ingest} we structure request handlers in a specific way and differentiate between root and host handlers. Figure~\ref{fig:nmi-http-request-handling} illustrates the request flow.

Since application handlers are always mapped to internal paths, Fiber will pass the request to root handlers in the beginning. First, \texttt{telemetry} gathers data about the request and afterwards the response (e.g.~client data or response times). Currently, this data is only provided in the form of logging, but exporting metrics would also be possible. Next, \texttt{request ID} will assign a random ID to this request that can help to trace and debug individual requests. Central to the \texttt{http} component is the \texttt{root router} handler that implements our custom routing logic. We will detail this process in the next paragraph. The \texttt{root router} will redirect the request internally to host handlers. If a set of host handlers does not handle the request, the \texttt{root router} might redirect to another matching set of host handlers. In case there is no more matching set of host handlers, the request will flow to the \texttt{catch-all deny} handler, that responds with an HTTP error code. Before handing the request to the application, there might be further host handlers. The \texttt{host route check} handler forbids direct access to internal routes and checks if there was an internal redirect beforehand. Next, \texttt{auth} and \texttt{CORS} handlers are inserted depending on the configuration of the application. Only after that, the request is handed off to the application.

\texttt{root router} primarily implements host-based routing based on the application's host patterns. For this, we map the host pattern to the internal path \texttt{/sys/internal/hosts/} followed by the SHA256 hash of the host pattern. The hash is used in the path in order to limit the character set to 0-9 and a-f. Based on our routing logic, we can then utilize internal redirects to direct the flow of requests. Fiber will continue to match the redirected request to appropriate host handlers or return to the \texttt{root router} for the next internal redirect. For the host-based routing, \texttt{root router} first checks for exact matches. For instance, assume there are applications with host patterns of \texttt{"primary.example.com"} and \texttt{"*.example.com"}. Requests with a host of primary.example.com will then be routed preferably to applications with the first pattern, the exact match. If the applications do not register handlers for the given HTTP verb and path, applications with the wildcard are considered. There might further be ambiguity when considering patterns with globbing. We resolve this by sorting the considered host patterns by length. For example, \texttt{"*.example.com"} is tried before \texttt{"**"}. The assumption is that longer matching patterns are ``closer'' to an exact match. An implementation of the \texttt{RootRouter} interface provides this routing logic and further allows registering host handlers with the \texttt{Host} method. We extracted these interfaces into a separate package because they can be used independently of the \texttt{http} server component types.

\section{Event Streaming}
\label{sec:event-streaming}

\begin{figure*}[t]
  \centering
  \includegraphics[width=\textwidth]{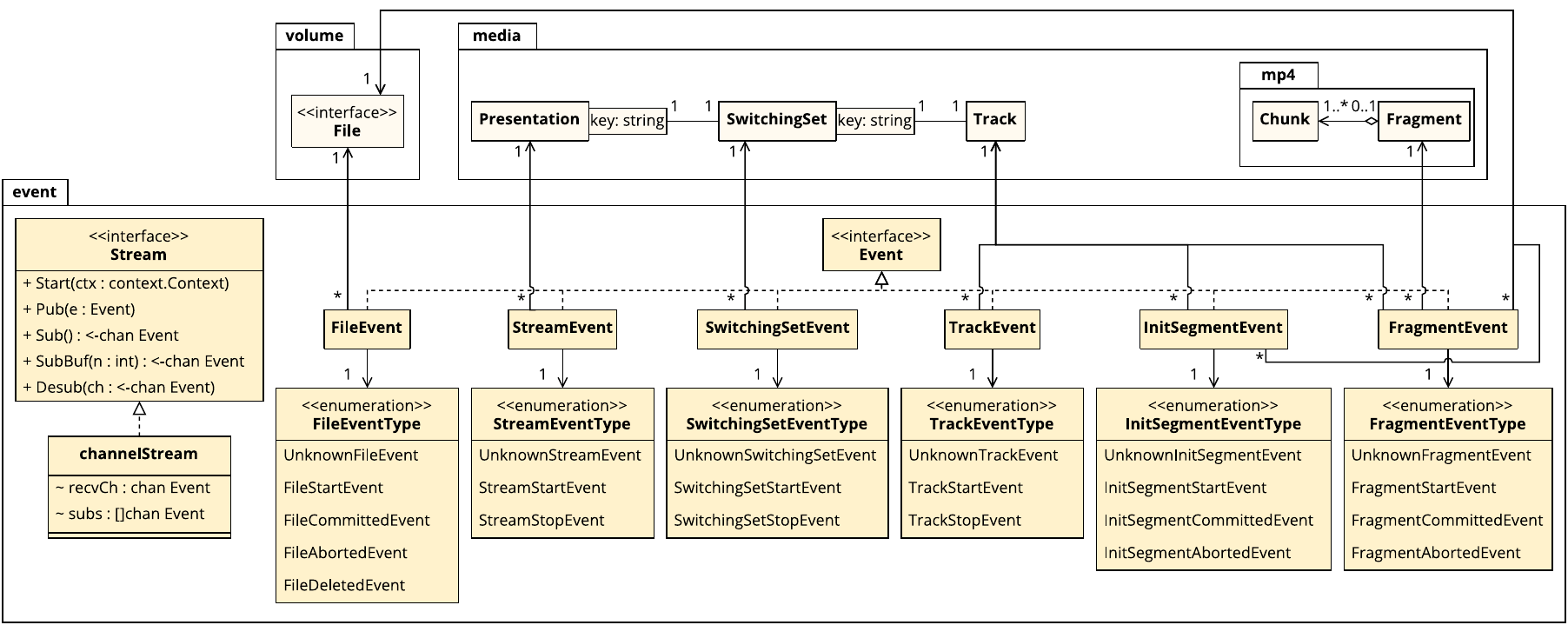}
  \caption{UML class diagram of the types in the \texttt{event} package.}
  \label{fig:nmi-uml-class-event}
  \Description{UML class diagram of the types in the \texttt{event} package.}
\end{figure*}

In this section, we will give more details about the event streaming implementation. As mentioned in the initial overview (see Section~\ref{sec:overview-of-nagare-media-ingest}), each application in \texttt{nagare media ingest} has an associated event stream. The application and the attached functions can publish events and subscribe to events in order to be notified. Figure~\ref{fig:nmi-uml-class-event} gives an overview of the relevant types.

The \texttt{Stream} interface describes the event streaming component. As such, it has a \texttt{Start} method to run it concurrently. \texttt{Pub} allows other components to publish \texttt{Event}s. \texttt{Sub} returns a receiving-only Go channel. From that point on, published \texttt{Event}s will be delivered to this channel. Each subscriber has its own Go channel. \texttt{Sub} returns a buffered channel with the default length of 32. New \texttt{Event}s are therefore delivered immediately as long as the buffer is not yet full. With unbuffered channels, the sender would block until the \texttt{Event} is received. If a subscriber is not directly available, e.g.~due to ongoing work with a previous \texttt{Event}, this might hinder the speedy delivery of \texttt{Events} to all other subscribers. Additionally, we allow subscribing with a buffered channel of custom length using the \texttt{SubBuf} method. The parameter \texttt{n} determines the length of the buffer; a length of 0 would return an unbuffered channel. Finally, the \texttt{Desub} method allows to unsubscribe and close the Go channel.

Various implementations are plausible, but we currently implement \texttt{Stream} only as \texttt{channelStream} that also uses Go channels internally and thus provides an easy in-memory solution. \texttt{Pub} will put the \texttt{Event} on the \texttt{recCh} channel. The event streaming Goroutine will receive it from there and fan-out the \texttt{Event} to all channels in the \texttt{subs} array.

\texttt{Event}s are typed. The \texttt{event} package already contains some \texttt{Event} types, but developers extending \texttt{nagare media ingest} can easily add new types. \texttt{FileEvent} should be used when a component operates on a \texttt{File} in a volume. It carries a reference to the corresponding \texttt{File} and the type of event that occurred as \texttt{FileEventType} enum, i.e.~the writing started, was committed, was aborted or the file was deleted. Components might additionally send more specific \texttt{Event}s in combination with \texttt{FileEvent}. For instance, \texttt{InitSegmentEvent} and \texttt{FragmentEvent} both also have an associated \texttt{File} reference, but that \texttt{File} has a specific role in the streaming protocol (see Subsection~\ref{subsec:cmafingest-dash-if-ingest-interface-1}). The last three provided \texttt{Event} types---\texttt{StreamEvent}, \texttt{SwitchingSetEvent} and \texttt{TrackEvent}---\linebreak{}describe start and end points of logical constructs of the DASH-IF ingest protocol as well as the CMAF format (see Subsection~\ref{subsec:cmafingest-dash-if-ingest-interface-1}). These \texttt{Event} types have corresponding references to relevant structures in the \texttt{media} package as well as an enum that describes the specific type of event.

\section{Application}
\label{sec:application}

Within the next subsections, we will outline our implementation of (ingest) applications. We again start with an overview in Subsection~\ref{subsec:application-overview}. Next, Subsections~\ref{subsec:cmafingest-dash-if-ingest-interface-1} and~\ref{subsec:dashandhlsingest-dash-if-ingest-interface-2} detail the implementation of the DASH-IF ingest protocol interface-1 and interface-2, respectively. Lastly, Subsection~\ref{subsec:genericserve} discusses a non-ingest application for retrieving ingested media.

\subsection{Overview}
\label{subsec:application-overview}

\begin{figure}[h]
  \centering
  \includegraphics[width=0.45\columnwidth]{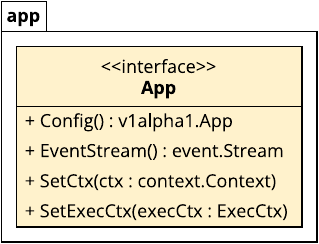}
  \caption{UML class diagram of the types in the \texttt{app} package.}
  \label{fig:nmi-uml-class-app}
  \Description{UML class diagram of the types in the \texttt{app} package.}
\end{figure}

Applications implement a certain ingest protocol. To encourage the composability and reuse of functionality, applications should only implement what is directly prescribed by the protocol, starting with the request handoff from the server until the written media files in a volume. Additional functionality should be extracted into concurrently running functions (see Section~\ref{sec:function}). An application should emit appropriate events to its event stream, giving other components the chance to couple additional logic to these events. Developers might add non-ingest applications for retrieval, monitoring or other purposes. Figure~\ref{fig:nmi-uml-class-app} depicts the application interface type.

The \texttt{App} interface is very generic. Depending on the server, applications might implement additional interfaces, for instance, \texttt{HTTPRegistrable} for \texttt{http} servers (see Subsection~\ref{subsec:http}). As such, the \texttt{App} interface only requires methods for accessing the application configuration as well as the event stream (\texttt{Config} and \linebreak{}\texttt{EventStream}). Moreover, \texttt{SetCtx} and \texttt{SetExecCtx} provide a way to initialize the application with the Go and \texttt{nagare media ingest} contexts, respectively.

\subsection{\texttt{cmafIngest}: DASH-IF Ingest Interface-1}
\label{subsec:cmafingest-dash-if-ingest-interface-1}
\begin{figure*}[t]
  \centering
  \includegraphics[width=0.9\textwidth]{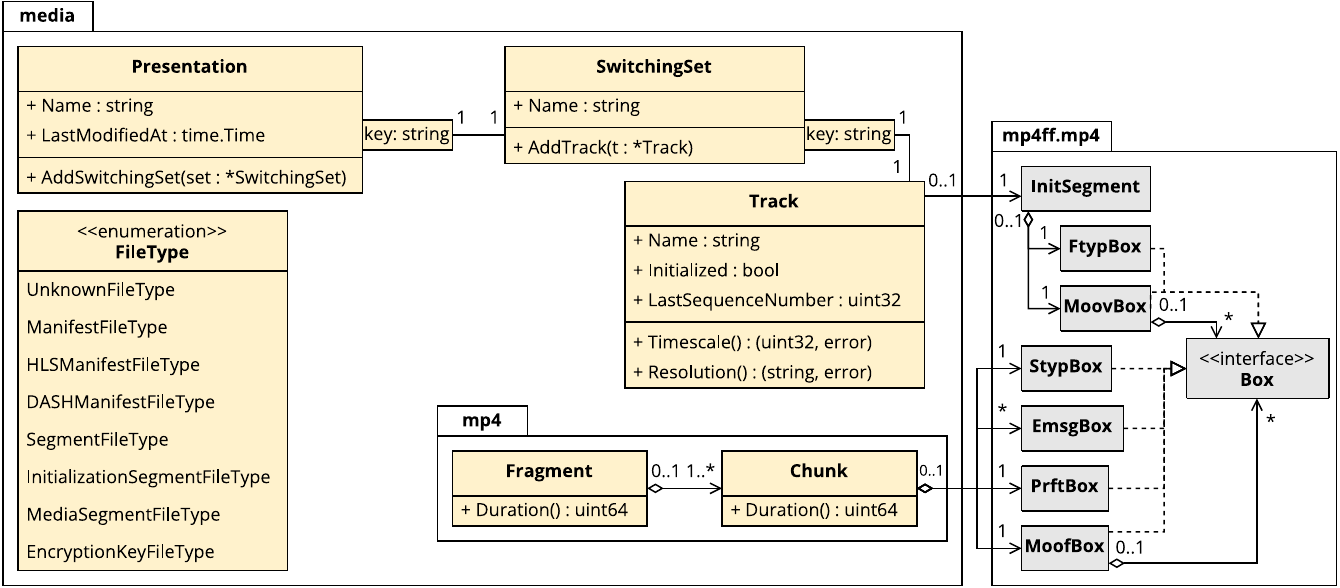}
  \caption{UML class diagram of the types in the \texttt{media} package.}
  \label{fig:nmi-uml-class-media}
  \Description{UML class diagram of the types in the \texttt{media} package.}
\end{figure*}

The \texttt{cmafIngest} application implements interface-1 of the \mbox{DASH-IF} ingest protocol also known as ``CMAF Ingest''. Standardized by MPEG in ISO/IEC~23000-19:2024~\cite{isoiec_isoiec2300019_2024}, CMAF is a common container format optimized for segmented media streaming. It builds upon the ISO~BMFF standard also published by MPEG~\cite{isoiec_isoiec1449612_2022}. CMAF, as a common format, allows reusing the same media objects for segmented media streaming protocols such as HLS and DASH. Thus, supporting multiple streaming protocols simultaneously no longer requires either storing multimedia files multiple times or special streaming servers that repackage multimedia streams on the fly. DASH-IF ingest interface-1 describes how CMAF media objects should be transmitted over HTTP/1.1 to the ingest server.

In ISO~BMFF, data is structured into boxes sometimes also called atoms. Each box has a specific type and size. Boxes can be structured hierarchically, i.e.~boxes can contain sub-boxes. This design allows easily extending the format by introducing new box types. Parsers should generally skip unknown boxes. Certain boxes are required in ISO~BMFF files. \texttt{ftyp}, the file type box, is typically the first box and describes to what brands, i.e.~specifications, this file best adheres to. For instance, CMAF defines the two brand identifiers \texttt{cmfc} and \texttt{cmf2} with corresponding constraints on multimedia files. Next, \texttt{moov}, the movie box, contains sub-boxes for specifying metadata. Together, \texttt{ftyp} and \texttt{moov} are considered to be the file header. Encoded media samples are stored in \texttt{mdat}, the media data box. For some applications, it can make sense to split the media samples over multiple boxes. This is mainly because \texttt{moov} sub-boxes will reference data in \texttt{mdat}. As such, the \texttt{moov} box can only be fully written after all media samples have been written to \texttt{mdat}. This is especially undesirable for multimedia streaming scenarios where the playback already starts during the creation of media samples. In these cases, media can be stored fragmented. Here, the \texttt{moov} box still contains relevant metadata, however, no (or not all) samples are referenced. Instead, \texttt{moof}, the movie fragment box, contains sub-boxes with the relevant fragment metadata. The duration of the media can thus be extended by concatenating \texttt{moof}-\texttt{mdat} pairs. The HLS and DASH streaming protocols use this feature and define segmented media files. Unlike ISO~BMFF files, segments do not contain an \texttt{ftyp} and \texttt{moov} box. Instead, it is recommended to start a segment with \texttt{styp}, the segment type box equivalent to \texttt{ftyp}. Segment files then at least contain one fragment, i.e.~one \texttt{moof} and \texttt{mdat} box.

In CMAF, media is always fragmented and additional notions are introduced. A \emph{CMAF header} consists of one \texttt{ftyp} and \texttt{moov} box. A \emph{CMAF chunk} is defined as exactly one \texttt{moof} and \texttt{mdat} pair\footnote{CMAF chunks can optionally further contain \texttt{styp}, \texttt{prft} and \texttt{emsg} boxes. For the remainder of this report, these do not play a significant role and are therefore omitted. \texttt{nagare media ingest} handles these boxes transparently, i.e.~they are ingested 1:1 without changes. Any other box type will result in an error.}. \emph{CMAF fragments} are then said to contain a list of chunks with the constraint that the first sample of the first chunk is a switching point. For newer MPEG video codecs, this constraint generally means that this sample is an instantaneous decoder refresh~(IDR) frame, a special kind of intra-coded frame~(I-frame) that signals to the decoder that decoding can be started independently of previous frames from this point on. As such, CMAF fragments (in combination with the CMAF header) can be decoded independently. This is required in multimedia streaming, e.g.~for players to join mid-stream or when seeking. \emph{CMAF segments} are defined as a list of CMAF fragments and are comparable to the segment definitions of the ISO~BMFF, HLS and DASH standards. Headers, chunks and segments are considered \emph{addressable media objects} that could be referenced in streaming protocols, e.g.~to be downloaded via HTTP. CMAF further defines logical structures that are not directly addressable. A \emph{CMAF track} logically consists of the CMAF header and a list of CMAF fragments. Tracks can be grouped into an \emph{(aligned) CMAF switching set}, for instance a video could be available in multiple resolutions and bit rates allowing players to switch between these tracks during playback. Switching sets can then be grouped into \emph{CMAF selection sets}, for example audio tracks could be available in different languages and the player selects the preferred one. Finally, a list of all CMAF selection sets is called a \emph{CMAF presentation}.

The DASH-IF ingest protocol uses these CMAF notions in order to define interface-1. However, noticeably absent are selection sets which are not mentioned in the standard. As such, we only defined Go structures corresponding to the other relevant CMAF notions. Figure~\ref{fig:nmi-uml-class-media} depicts the types in the \texttt{media} package. Note that we only modelled metadata fields; an association to an \texttt{mdat} box is therefore missing. Where possible, types from the \texttt{github.com/Eyevinn/mp4ff}~(mp4ff) library were reused.

\begin{figure}[h]
  \centering
  \includegraphics[width=0.75\columnwidth]{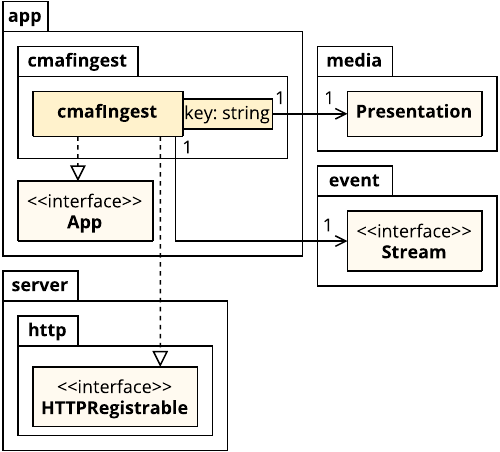}
  \caption{UML class diagram of the types in the \texttt{app.cmafingest} package.}
  \label{fig:nmi-uml-class-app-cmafingest}
  \Description{UML class diagram of the types in the \texttt{app.cmafingest} package.}
\end{figure}

In interface-1, CMAF media objects are ingested via HTTP/1.1 \texttt{POST} or \texttt{PUT} requests. Clients may issue requests for individual CMAF segments or ingest complete CMAF tracks using long-\linebreak{}running requests. In the latter case, HTTP/1.1 chunked transfer encoding~(CTE)~\cite{fielding_http11_2022} should be used to transfer the CMAF track in parts as it is being produced. \texttt{cmafIngest} currently only implements long-running requests. We define all media objects to belong to the same CMAF presentation if they share a common HTTP path prefix. For example, requests to \texttt{/app/example.str/\linebreak{}Switching(video)/Stream(1080p.cmfv)} and \texttt{/app/example\linebreak{}.str/Switching(audio)/Stream(en.cmfa)} would ingest to the same CMAF presentation ``example''. Interface-1 defines multiple ways to signal to what track and switching sets media objects belong to. The \texttt{Switching()} and \texttt{Stream()} keywords provide a direct way in the HTTP path to set the switching set and track, respectively. Alternatively, clients can first ingest a DASH or HLS manifest that defines switching sets and a naming structure that is then followed by later requests. Finally, clients could add a \texttt{kind} metadata box to the media object that contains the switching set ID. \texttt{cmafIngest} currently only implements direct signaling through the \texttt{Switching()} and \texttt{Stream()} keywords and leaves the other options for future work. Figure~\ref{fig:nmi-uml-class-app-cmafingest} shows the types we implemented in the \texttt{app.cmafingest} package.

\texttt{cmafIngest} writes CMAF headers and CMAF fragments as individual files to the configured volume. Moreover, appropriate events are emitted. For fragments, we use an in-place writer to enable low-latency use cases. We decode the \texttt{moof} box of each CMAF chunk in order to determine the fragment boundaries, i.e.~where the first media sample is a switching point. Note that we only decode the metadata and not the media itself. As such, this process is quick. Next to the volume, administrators can further configure size limits for CMAF objects to combat misuse or protect against a faulty client. Furthermore, a timeout determines when a CMAF presentation is considered to be terminated after a sudden interruption of the ingest. \texttt{cmafIngest} will clean up in-memory representations of terminated CMAF presentations regularly.

\subsection{\texttt{dashAndHlsIngest}: DASH-IF Ingest Interface-2}
\label{subsec:dashandhlsingest-dash-if-ingest-interface-2}

Next to interface-1, DASH-IF ingest also defines interface-2, the ``DASH and HLS Ingest'' protocol. The media may already be packaged by the client for delivery via DASH or HLS. In this case, the client could use interface-2 to transfer the packaged presentation to the ingest server. Interface-2 is simpler, as the ingest server does not need to decode the incoming data and can assume the packaging is correct. Unlike interface-1, where the DASH or HLS manifests were optionally used to signal switching sets and tracks, interface-2 requires the transmission of the whole presentation including manifests. This also means that manifest updates, e.g.~after appending media segments, necessitate repeated ingests. It is still recommended to conform the media to the CMAF standard, but alternative formats supported by DASH and HLS can be used, too. As in interface-1, Clients send HTTP/1.1 \texttt{POST} or \texttt{PUT} requests to ingest data. HTTP CTE for low latency is also possible. Additionally, \texttt{DELETE} requests are supported to remove previously ingested data. This is useful for sliding window streaming, where at any given time only a limited number of segments are available, i.e.~as new segments are appended to the presentation, the oldest segments are deleted. In this way, server resources are conserved. Interface-2 thus defines a common specification for DASH and HLS ingests of what have previously only been industry best practices.

\begin{figure}[h]
  \centering
  \includegraphics[width=0.8\columnwidth]{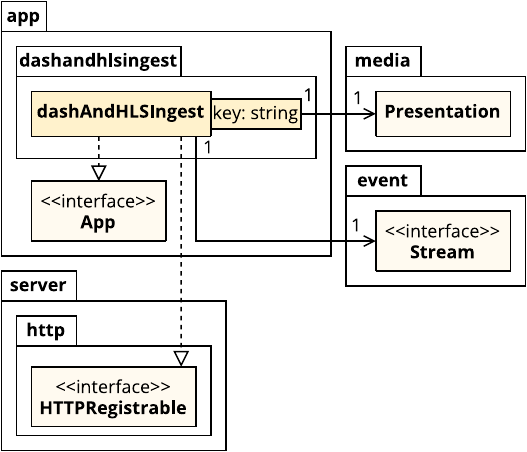}
  \caption{UML class diagram of the types in the \texttt{app.dashandhlsingest} package.}
  \label{fig:nmi-uml-class-app-dashandhlsingest}
  \Description{UML class diagram of the types in the \texttt{app.dashandhlsingest} package.}
\end{figure}

We implemented interface-2 in the \texttt{dashAndHLSIngest} application. As with \texttt{cmafIngest}, we regard ingested data to be part of the same presentation if they share a common path prefix (e.g.~\texttt{/app/\linebreak{}example.str/index.m3u8} and \texttt{/app/example.str/video/720p-\linebreak{}07.cmfv}). Administrators can choose if in-place writers should be used. They can further set size limits for ingested files. Figure~\ref{fig:nmi-uml-class-app-dashandhlsingest} depicts the types of our implementation.

\subsection{\texttt{genericServe}}
\label{subsec:genericserve}

\begin{figure}[h]
  \centering
  \includegraphics[width=0.7\columnwidth]{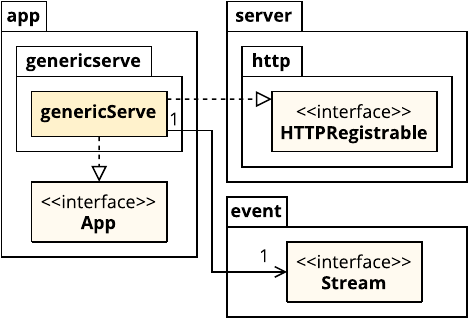}
  \caption{UML class diagram of the types in the \texttt{app.genericserve} package.}
  \label{fig:nmi-uml-class-app-genericserve}
  \Description{UML class diagram of the types in the \texttt{app.genericserve} package.}
\end{figure}

To be useful, \texttt{nagare media ingest} must provide ways for integrating with other systems. \texttt{genericServe} is a non-ingest HTTP application for retrieving previously ingested data. Administrators can configure \texttt{genericServe} alongside an ingest application and thus realize use cases with external systems, e.g.~a workflow system that retrieves and further processes multimedia data. A \texttt{genericServe} instance references exactly one ingest application as well as a list of volumes. It handles \texttt{GET} requests by mapping the request path to the files of the referenced application. For instance, requesting \texttt{/app/example.str/video/720p-07.cmfv} would search for the file \texttt{/ref-app/example.str/video/720p-07.cmfv}. For this, \texttt{genericServe} iterates over each volume and returns the first match. If none of the volumes has this file, it responds with a \texttt{Not Found} error. The list of volumes thus forms one logical filesystem that allows for advanced configurations. For example, the \texttt{cmafIngest} application could ingest CMAF media objects to an appropriately sized \texttt{fs} volume. Simultaneously, a connected \texttt{manifest} function (see Subsection~\ref{subsec:manifest}) could use a \texttt{mem} volume to store frequently accessed and updated HLS manifests. The \texttt{genericServe} application is then able to read files from both volumes with the corresponding benefits (size versus speed). Figure~\ref{fig:nmi-uml-class-app-genericserve} depicts the relevant types and associations for \texttt{genericServe}.

We make sure to handle and set appropriate HTTP headers for file delivery. Based on the file extension, the \texttt{Content-Type} is set. Administrators can configure a default in case the type cannot be determined (e.g.~\texttt{application/octet-stream} as generic type). The \texttt{Last-Modified} header is set for all volume types (cf.~\texttt{ModTime} method in Section~\ref{sec:volume}). \texttt{genericServe} supports retrieval of files with in-place writers. HTTP/1.1 CTE is used to deliver these files in parts as they become available. In this case, the \texttt{Transfer-Encoding} header is set to \texttt{chunked}. In contrast, the \texttt{Content-Length} and \texttt{ETag} headers are set for files without an in-place writer. Moreover, range requests are supported for complete files. Here, HTTP servers can indicate they support range requests using the \texttt{Accept-Ranges} response header and HTTP clients can request partial files using the \texttt{Content-Range} request header. \texttt{genericServe} then only returns the requested byte range in a partial content response. This is often used in multimedia streaming, e.g.~when seeking to a particular point in a media object. For the particular scenario of running \texttt{nagare media ingest} behind a reverse proxy, \texttt{genericServe} supports the non-standard headers \texttt{X-Sendfile} and \texttt{X-Accel-Redirect} for the web servers Apache httpd\footnote{\url{https://httpd.apache.org/}} and nginx\footnote{\url{https://nginx.org/}}, respectively. If enabled in the configuration, \texttt{genericServe} will only return an empty response with these headers containing the path to the requested file. The reverse proxy will then recognize these headers and replace the empty response with the content of the file. This effectively offloads the responsibility of serving files to the reverse proxy. Of course, this is only possible with \texttt{fs} volumes. But as mentioned in Subsection~\ref{subsec:fs}, we carefully implemented the \texttt{fs} volume to make use of the \texttt{sendfile} UNIX system call so serving files should be performant even without the usage of highly optimized reverse proxies.

\section{Function}
\label{sec:function}

Functions are the last major component type in \texttt{nagare media ingest}. In Subsection~\ref{subsec:function-overview} of this section, we discuss this component in more detail. This is followed by concrete function implementations in Subsections~\ref{subsec:copy} to~\ref{subsec:cleanup}. These are currently available in \texttt{nagare media ingest} and can help administrators with specific use cases. They can also serve as examples for developers that want to implement custom functions.

\subsection{Overview}
\label{subsec:function-overview}

\begin{figure}[h]
  \centering
  \includegraphics[width=0.75\columnwidth]{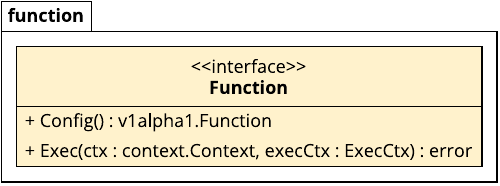}
  \caption{UML class diagram of the types in the \texttt{function} package.}
  \label{fig:nmi-uml-class-function}
  \Description{UML class diagram of the types in the \texttt{function} package.}
\end{figure}

Function instances are bound to one ingest application and extend the ingest protocol in order to implement a specific requirement. Ideally, functions are implemented in a generic way such that they can be used with various ingest protocols. Figure~\ref{fig:nmi-uml-class-function} depicts the \texttt{Function} interface.

The \texttt{Config} method returns the configuration passed to the function during initialization. Each function instance runs concurrently in its own Goroutine created by the function controller (see Subsection~\ref{subsec:implementation}). Developers can implement the logic of the \texttt{Exec} method in any way, but we expect most functions to follow an event loop approach. The passed \texttt{ExecCtx} structure allows access to the context of the function, i.e.~the application the function is bound to as well as existing volumes. We again use the Go \texttt{Context} to propagate termination requests. Developers should therefore make sure to handle canceled Go \texttt{Context}s. Through \texttt{ExecCtx}, functions can subscribe and publish to the application's event stream. Listing~\ref{lst:function} demonstrates how an event loop can be structured in a minimal example function.
\begin{listing}
\begin{minted}{go}
type example struct {
  cfg v1alpha1.Function
}

func New(cfg v1alpha1.Function) (Function, error) {
  return &example{cfg: cfg}, nil
}

func (f *example) Config() v1alpha1.Function {
  return f.cfg
}

func (f *example) Exec(ctx context.Context,
    execCtx ExecCtx) error {
  l := execCtx.Logger()
  es := execCtx.App().EventStream().Sub()
  defer execCtx.App().EventStream().Desub(es)

  for {
    select {
    case <-ctx.Done():
      l.Info("terminate example function")
      return nil

    case ee := <-es:
      switch e := ee.(type) {
      case *event.StreamEvent:
        l.Infof(
          "CMAF Presentation event for %s",
          e.Stream.Name
        )
      case *event.SwitchingSetEvent:
        l.Infof(
          "CMAF Switching Set event for %s",
          e.SwitchingSet.Name
        )
      case *event.TrackEvent:
        l.Infof(
          "CMAF Track event for %s",
          e.Track.Name
        )
      default:
        l.Infof("unhandled event of type %T", e)
      }
    }
  }
}
\end{minted}
  \caption{Example function implementation.}
  \label{lst:function}
  \Description{Example function implementation.}
\end{listing}

Here, no configuration options are relevant. The constructor method thus just initializes a new \texttt{example} structure and passes the configuration to the \texttt{cfg} field (lines~1--7). The \texttt{Config} method returns this field accordingly (lines~9--11). In \texttt{Exec} (lines~13--47), we first retrieve the logger (line~15) and subscribe to the application event stream (line~16). Go's \texttt{defer} statement is used to unsubscribe from the event stream once the Goroutine returns from \texttt{Exec} (line~17). The following lines then define the event loop. Go's \texttt{select} statement will pause the Goroutine until either the Go \texttt{Context} was canceled (line~21) or a new event arrives (line~25). In the first case, we log a termination message and return from \texttt{Exec} (lines~22 and~23). When instead a new event arrives, we use Go's type switch statement in order to handle various CMAF event types (lines~26--44; also see Section~\ref{sec:event-streaming}). Note that the variable \texttt{e} is correctly typed in each \texttt{case}, allowing us to access event-specific fields when logging messages. The \texttt{default} case with a generic log message will be executed if none of the listed event types matches (lines~42--44). All the following functions are structured in this way.

\subsection{\texttt{copy}}
\label{subsec:copy}

\begin{figure}[h]
  \centering
  \includegraphics[width=0.6\columnwidth]{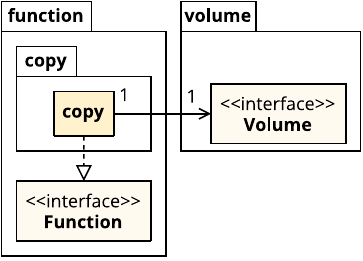}
  \caption{UML class diagram of the types in the \texttt{function.copy} package.}
  \label{fig:nmi-uml-class-function-copy}
  \Description{UML class diagram of the types in the \texttt{function.copy} package.}
\end{figure}

The \texttt{copy} function copies files from one volume to another. It can be applied, for instance, for archiving purposes. Another example use case would be to temporarily ingest first to a fast \texttt{mem} volume and simultaneously copy to an \texttt{fs} volume for long-term storage. Figure~\ref{fig:nmi-uml-class-function-copy} depicts our implementation.

\texttt{copy} is structured around an event loop that listens for \texttt{File\linebreak{}Event}s of type \texttt{FileCommittedEvent} (see Section~\ref{sec:event-streaming}). These events carry a reference to a \texttt{File} object from which \texttt{copy} acquires a reader. It then opens a \texttt{File} under the same path in the configured volume for writing. Finally, data from the reader is copied over to the writer. For each copy process a separate Goroutine is created such that simultaneous copies are possible. When the Go \texttt{Context} is canceled, we make sure to delay the termination until all copy operations are finished.

\subsection{\texttt{manifest}}
\label{subsec:manifest}

\begin{figure}[h]
  \centering
  \includegraphics[width=\columnwidth]{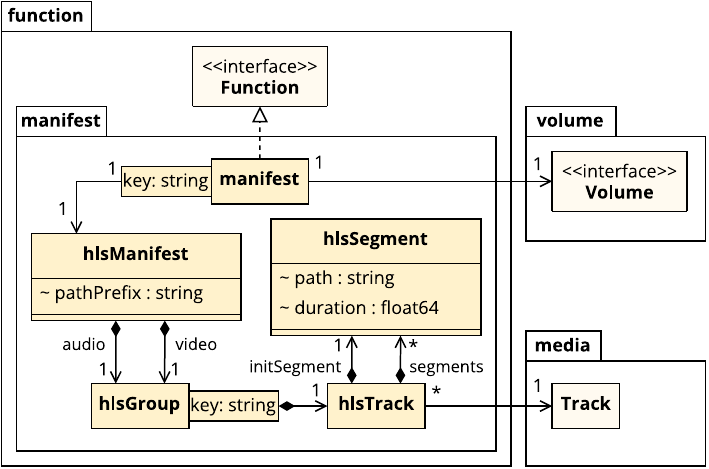}
  \caption{UML class diagram of the types in the \texttt{function.manifest} package.}
  \label{fig:nmi-uml-class-function-manifest}
  \Description{UML class diagram of the types in the \texttt{function.manifest} package.}
\end{figure}

The DASH and HLS multimedia streaming protocols define manifest formats that index and reference multimedia objects, e.g.~CMAF segments. Ingesting a CMAF presentation with the \texttt{cmafIngest} application (see Subsection~\ref{subsec:cmafingest-dash-if-ingest-interface-1}) still lacks manifest files and is therefore not directly playable. The \texttt{manifest} function generates HLS manifests based on the ingested CMAF presentation and stores them in a configured volume. The required types of our implementation are shown in Figure~\ref{fig:nmi-uml-class-function-manifest}.

CMAF notions (see Subsection~\ref{subsec:cmafingest-dash-if-ingest-interface-1}) can be mapped to HLS. For this, we defined various types with the necessary fields to generate HLS manifests. Each CMAF presentation will result in an instance of \texttt{hlsManifest}. CMAF switching sets are mapped to \texttt{hlsGroup}, CMAF tracks to \texttt{hlsTrack} and CMAF segments/fragments to \linebreak{}\texttt{hlsSegment} objects.

The \texttt{manifest} function listens for \texttt{InitSegmentEvent}s and \linebreak{}\texttt{FragmentEvent}s of type \texttt{Committed} (see Section~\ref{sec:event-streaming}). \texttt{InitSegment\linebreak{}Event}s have an associated \texttt{Track} object that points to an \linebreak{}\texttt{InitSegment}, i.e.~the CMAF header. Based on this metadata, we categorize \texttt{Track}s in an audio or video \texttt{hlsGroup}\footnote{Note that HLS allows for more groups including alternative audio/video groups or groups for subtitles. We simplified the implementation for our prototype.}. We further store the path to the CMAF header by referring to the \texttt{File} object referenced in the event. \texttt{FragmentEvent}s will update the list of \texttt{hlsSegment}s of the corresponding \texttt{hlsTrack}. After each event, we regenerate the changed HLS manifest, i.e.~the multivariant manifest after a new CMAF header and the media manifest after segment updates. Generated manifests are written to the configured volume.

\subsection{\texttt{cloudEvent}}
\label{subsec:cloudevent}

\begin{figure}[h]
  \centering
  \includegraphics[width=0.55\columnwidth]{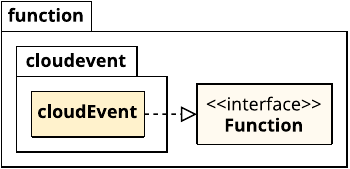}
  \caption{UML class diagram of the types in the \texttt{function.cloudevent} package.}
  \label{fig:nmi-uml-class-function-cloudevent}
  \Description{UML class diagram of the types in the \texttt{function.cloudevent} package.}
\end{figure}

In order to facilitate integrations with external software (e.g.~a workflow system), the \texttt{cloudEvent} function provides a way to send encoded events to an HTTP endpoint. This approach is often referred to as webhook. Events could be encoded in various ways. We choose the CloudEvents standard that is maintained by the Cloud Native Computing Foundation~(CNCF) and defines a generic format often serialized as JSON object~\cite{cncf_cloudeventsversion102_2022}. All CloudEvents have common fields such as \texttt{type}, \texttt{id}, \texttt{time}, \texttt{source} or \texttt{subject}, but can additionally carry custom data. We use the official Go library \texttt{github.com/cloudevents/sdk-go/v2}\footnote{\url{https://github.com/cloudevents/sdk-go}} to format CloudEvents. Figure~\ref{fig:nmi-uml-class-function-cloudevent} depicts the \texttt{cloudEvent} function implementation.

The event loop of \texttt{cloudEvent} creates a CloudEvent and sends it to a configured URL. \texttt{nagare media ingest} events are embedded as serialized JSON objects in CloudEvents. Developers defining event structures can therefore employ Go's struct tags as the typical way to influence the serialization process.

\subsection{\texttt{cleanup}}
\label{subsec:cleanup}

\begin{figure}[h]
  \centering
  \includegraphics[width=0.6\columnwidth]{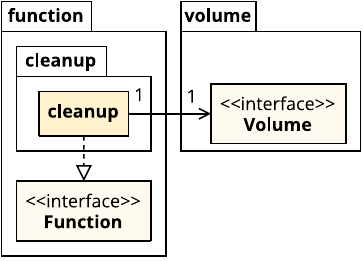}
  \caption{UML class diagram of the types in the \texttt{function.cleanup} package.}
  \label{fig:nmi-uml-class-function-cleanup}
  \Description{UML class diagram of the types in the \texttt{function.cleanup} package.}
\end{figure}

The last function is \texttt{cleanup}. As the name suggests, it removes certain files from volumes. Currently, administrators can configure file patterns and a certain age. If any of the patterns matches the file path and if the file exceeds the specified age, it will be deleted from all configured volumes. As such, this function can be useful when implementing sliding window streaming where old media objects are deleted as new ones are ingested. For pattern matching, we again use the \texttt{github.com/gobwas/glob} library for UNIX-like globbing (cf. Subsection~\ref{subsec:http}). Future work could extend this function with additional matchers. In Figure~\ref{fig:nmi-uml-class-function-cleanup}, the necessary types are depicted.

\texttt{cleanup} listens for \texttt{FileEvent}s of type \texttt{FileCommittedEvent} and then checks whether the written file matches one of the configured patterns. If that is the case, it will be added to the end of a linked list. Because events arrive in order, this list will be sorted by file age. A second Goroutine repeatedly retrieves the age of the first element in the list and waits until the configured time. After waking up, it goes through the list and deletes files until it finds a file that is not yet of the defined age. It will then sleep again and repeat that process. In case the list is empty, it will sleep for a default time of 10~seconds. When deleting a file, \texttt{cleanup} emits a \texttt{FileEvent} of type \texttt{FileDeletedEvent} to the application's event stream.

\section{Conclusion}
\label{sec:conclusion}

This section concludes the technical report. Hence, Subsection~\ref{subsec:summary} summarizes our work on \texttt{nagare media ingest}. Finally, Subsection~\ref{subsec:future-work} shows potentials for future work.

\subsection{Summary}
\label{subsec:summary}

This technical report outlined the open source software \texttt{nagare media ingest}. It implements a system for modern multimedia ingest workflows in cloud and edge environments. A high degree of flexibility is achieved due to its design allowing the implementation of numerous ingest protocols and use cases. Functionality is split into various components that are instantiated by administrators through configuration. Volumes provide a file-based interface for storing data. Servers provide network-related logic for listening and accepting ingest requests. Applications are deployed on top of servers in order to handle requests and thus implement concrete ingest protocols. Functions run concurrently to an application and extended the base-functionality of the ingest protocol, hence cater for a particular use case. The application and associated functions do not communicate directly but instead send events to an application event stream. Functions therefore mainly take the form of an event loop that listens for and reacts to certain events. Depending on the type, events can carry references to additional in-memory structures relevant to the event.

In our prototype, we implement concrete examples for each component type. With \texttt{null}, a dummy volume implementation is provided for testing purposes. \texttt{mem} and \texttt{fs} volumes store files in-memory and in the local filesystem, respectively. The \texttt{http} server provides an environment for HTTP/1.1 applications. Next, the \texttt{cmafIngest} and \texttt{dashAndHlsIngest} applications implement interface-1 (CMAF Ingest) and interface-2 (DASH and HLS Ingest) of the DASH-IF ingest protocol. Additionally, the non-ingest application \texttt{genericServe} allows retrieving previously ingested files via HTTP. Further use cases are enabled with a set of functions. \texttt{copy} duplicates files to another volume. \texttt{manifest} creates HLS manifests appropriate for immediate playback of ingested CMAF media objects. \texttt{cloudevent} forwards arriving events to an HTTP webhook following the CloudEvents specification. Lastly, \texttt{cleanup} deletes files that reach a configured age, e.g.~for sliding window streaming.

\subsection{Future Work}
\label{subsec:future-work}

\texttt{nagare media ingest} has a design that lends itself to extensions. Additional volumes, servers, applications and functions could be implemented in future iterations. We would like to see support for more of the modern ingest protocols such as MOQT. Next to live content, VoD ingests could be added by implementing protocols of popular media asset management systems. Similarly, further functions would enhance \texttt{nagare media ingest} and open it up for more ingest workflows. For instance, an archive function could generate CMAF track files optimized for archival purposes. Moreover, an S3 volume implementation would improve cloud and edge integrations.

At the same time, some of the existing components can be improved. Currently, \texttt{cmafIngest} does not support all signaling options for CMAF switching sets and tracks. Extending support would lead to a full DASH-IF ingest implementation. Additionally, the implemented ingest applications could be more forgiving with non-standard clients, leading to an increased fault tolerance. Furthermore, the garbage collection processes of the \texttt{mem} and \texttt{fs} volumes would benefit from performance optimizations for high-contention situations. Finally, \texttt{cleanup} would serve more use cases with additional matchers.

Future work could also tackle the base functionality of \texttt{nagare media ingest}. The event system currently published events to every subscriber. Administrators might want to influence this behavior, e.g.~by setting up filters or routes. Running \texttt{nagare media ingest} in a production setting necessitates increased observability. Therefore, gathering and providing telemetry data, e.g.~in the form of metrics, logging or request tracing, would make the system more production-ready. On a similar note, future work could investigate the behavior of \texttt{nagare media ingest} in various production use cases to validate its benefits and identify limitations.

%% file: main.bbl

\begin{thebibliography}{24}


\ifx \showCODEN    \undefined \def \showCODEN     #1{\unskip}     \fi
\ifx \showISBNx    \undefined \def \showISBNx     #1{\unskip}     \fi
\ifx \showISBNxiii \undefined \def \showISBNxiii  #1{\unskip}     \fi
\ifx \showISSN     \undefined \def \showISSN      #1{\unskip}     \fi
\ifx \showLCCN     \undefined \def \showLCCN      #1{\unskip}     \fi
\ifx \shownote     \undefined \def \shownote      #1{#1}          \fi
\ifx \showarticletitle \undefined \def \showarticletitle #1{#1}   \fi
\ifx \showURL      \undefined \def \showURL       {\relax}        \fi
\providecommand\bibfield[2]{#2}
\providecommand\bibinfo[2]{#2}
\providecommand\natexlab[1]{#1}
\providecommand\showeprint[2][]{arXiv:#2}

\bibitem[{Aguilar-Armijo} et~al\mbox{.}(2020)]%
        {aguilararmijo_dynamicsegmentrepackaging_2020}
\bibfield{author}{\bibinfo{person}{Jesus {Aguilar-Armijo}}, \bibinfo{person}{Babak Taraghi}, \bibinfo{person}{Christian Timmerer}, {and} \bibinfo{person}{Hermann Hellwagner}.} \bibinfo{year}{2020}\natexlab{}.
\newblock \showarticletitle{Dynamic {{Segment Repackaging}} at the {{Edge}} for {{HTTP Adaptive Streaming}}}. In \bibinfo{booktitle}{\emph{2020 {{IEEE International Symposium}} on {{Multimedia}} ({{ISM}})}}. \bibinfo{publisher}{IEEE}, \bibinfo{address}{Naples, Italy}, \bibinfo{pages}{17--24}.
\newblock
\showISBNx{978-1-7281-8697-9}
\href{https://doi.org/10.1109/ISM.2020.00009}{doi:\nolinkurl{10.1109/ISM.2020.00009}}


\bibitem[{Ben-Kiki} et~al\mbox{.}(2004)]%
        {benkiki_yamlaintmarkup_2004}
\bibfield{author}{\bibinfo{person}{Oren {Ben-Kiki}}, \bibinfo{person}{Clark Evans}, {and} \bibinfo{person}{Brian Ingerson}.} \bibinfo{year}{2004}\natexlab{}.
\newblock \bibinfo{booktitle}{\emph{{{YAML Ain}}'t {{Markup Language}} ({{YAML}}™) 1.0}}.
\newblock \bibinfo{type}{Standard}.
\newblock
\urldef\tempurl%
\url{https://yaml.org/spec/1.0/}
\showURL{%
Retrieved 2025-09-08 from \tempurl}


\bibitem[{Bitmovin Inc.}(2025)]%
        {bitmovininc_8thannualbitmovin_2025}
\bibfield{author}{\bibinfo{person}{{Bitmovin Inc.}}} \bibinfo{year}{2025}\natexlab{}.
\newblock \bibinfo{title}{The 8th {{Annual Bitmovin Video Developer Report}}}.
\newblock
\urldef\tempurl%
\url{https://bitmovin.com/video-developer-report/}
\showURL{%
Retrieved 2025-05-21 from \tempurl}


\bibitem[CNCF(2022)]%
        {cncf_cloudeventsversion102_2022}
\bibfield{author}{\bibinfo{person}{CNCF}.} \bibinfo{year}{2022}\natexlab{}.
\newblock \bibinfo{booktitle}{\emph{{{CloudEvents}} - {{Version}} 1.0.2}}.
\newblock \bibinfo{type}{{T}echnical {R}eport}. \bibinfo{institution}{Cloud Native Computing Foundation}.
\newblock
\urldef\tempurl%
\url{https://github.com/cloudevents/spec/blob/v1.0.2/cloudevents/spec.md}
\showURL{%
Retrieved 2022-02-24 from \tempurl}


\bibitem[{DASH-IF}(2024)]%
        {dashif_dashiflivemedia_2024}
\bibfield{author}{\bibinfo{person}{{DASH-IF}}.} \bibinfo{year}{2024}\natexlab{}.
\newblock \bibinfo{booktitle}{\emph{{{DASH-IF Live Media Ingest Protocol}} 1.2}}.
\newblock \bibinfo{type}{{T}echnical {R}eport}. \bibinfo{institution}{DASH Industry Forum}.
\newblock
\urldef\tempurl%
\url{https://dashif.org/Ingest/}
\showURL{%
Retrieved 2025-05-13 from \tempurl}


\bibitem[ETSI(2025)]%
        {etsi_etsits126_2025}
\bibfield{author}{\bibinfo{person}{ETSI}.} \bibinfo{year}{2025}\natexlab{}.
\newblock \bibinfo{booktitle}{\emph{{{ETSI TS}} 126 512 {{5G}}; {{5G Media Streaming}} ({{5GMS}}); {{Protocols}} ({{3GPP TS}} 26.512 version 18.5.0 {{Release}} 18)}}.
\newblock \bibinfo{type}{Standard}. \bibinfo{institution}{European Telecommunications Standards Institute}, \bibinfo{address}{Sophia Antipolis, FR}.
\newblock


\bibitem[Fielding et~al\mbox{.}(2022)]%
        {fielding_http11_2022}
\bibfield{author}{\bibinfo{person}{Roy~T. Fielding}, \bibinfo{person}{Mark Nottingham}, {and} \bibinfo{person}{Julian Reschke}.} \bibinfo{year}{2022}\natexlab{}.
\newblock \bibinfo{title}{{{HTTP}}/1.1}.
\newblock \bibinfo{howpublished}{RFC 9112}.
\newblock
\href{https://doi.org/10.17487/RFC9112}{doi:\nolinkurl{10.17487/RFC9112}}


\bibitem[ISO/IEC(2022a)]%
        {isoiec_isoiec1449612_2022}
\bibfield{author}{\bibinfo{person}{ISO/IEC}.} \bibinfo{year}{2022}\natexlab{a}.
\newblock \bibinfo{booktitle}{\emph{{{ISO}}/{{IEC}} 14496-12:2022 {{Information}} technology -- {{Coding}} of audio-visual objects -- {{Part}} 12: {{ISO}} base media file format}}.
\newblock \bibinfo{type}{Standard}. \bibinfo{institution}{International Organization for Standardization}, \bibinfo{address}{Geneva, CH}.
\newblock


\bibitem[ISO/IEC(2022b)]%
        {isoiec_isoiec230091_2022}
\bibfield{author}{\bibinfo{person}{ISO/IEC}.} \bibinfo{year}{2022}\natexlab{b}.
\newblock \bibinfo{booktitle}{\emph{{{ISO}}/{{IEC}} 23009-1:2022 {{Information}} technology -- {{Dynamic}} adaptive streaming over {{HTTP}} ({{DASH}}) -- {{Part}} 1: {{Media}} presentation description and segment formats}}.
\newblock \bibinfo{type}{Standard}. \bibinfo{institution}{International Organization for Standardization}, \bibinfo{address}{Geneva, CH}.
\newblock


\bibitem[ISO/IEC(2023)]%
        {isoiec_recituth2220_2023}
\bibfield{author}{\bibinfo{person}{ISO/IEC}.} \bibinfo{year}{2023}\natexlab{}.
\newblock \bibinfo{booktitle}{\emph{Rec. {{ITU-T H}}.222.0 v9 (08/23) {\textbar} {{ISO}}/{{IEC}} 13818-1:2023 {{Information}} technology -- {{Generic}} coding of moving pictures and associated audio information -- {{Part}} 1: {{Systems}}}}.
\newblock \bibinfo{type}{Standard}. \bibinfo{institution}{International Organization for Standardization}, \bibinfo{address}{Geneva, CH}.
\newblock


\bibitem[ISO/IEC(2024)]%
        {isoiec_isoiec2300019_2024}
\bibfield{author}{\bibinfo{person}{ISO/IEC}.} \bibinfo{year}{2024}\natexlab{}.
\newblock \bibinfo{booktitle}{\emph{{{ISO}}/{{IEC}} 23000-19:2024 {{Information}} technology -- {{Multimedia}} application format ({{MPEG-A}}) -- {{Part}} 19: {{Common}} media application format ({{CMAF}}) for segmented media}}.
\newblock \bibinfo{type}{Standard}. \bibinfo{institution}{International Organization for Standardization}, \bibinfo{address}{Geneva, CH}.
\newblock


\bibitem[{Jamie Fletcher}(2020)]%
        {jamiefletcher_livemediaingest_2020}
\bibfield{author}{\bibinfo{person}{{Jamie Fletcher}}.} \bibinfo{year}{2020}\natexlab{}.
\newblock \bibinfo{title}{Live {{Media Ingest}} ({{CMAF}})}.
\newblock
\urldef\tempurl%
\url{https://www.unified-streaming.com/blog/live-media-ingest-cmaf}
\showURL{%
Retrieved 2022-02-24 from \tempurl}


\bibitem[Mekuria et~al\mbox{.}(2020)]%
        {mekuria_toolslivecmaf_2020}
\bibfield{author}{\bibinfo{person}{Rufael Mekuria}, \bibinfo{person}{Dirk Griffioen}, {and} \bibinfo{person}{Arjen Wagenaar}.} \bibinfo{year}{2020}\natexlab{}.
\newblock \showarticletitle{Tools for live {{CMAF}} ingest}. In \bibinfo{booktitle}{\emph{Proceedings of the 11th {{ACM Multimedia Systems Conference}}}}. \bibinfo{publisher}{ACM}, \bibinfo{address}{Istanbul Turkey}, \bibinfo{pages}{273--278}.
\newblock
\showISBNx{978-1-4503-6845-2}
\href{https://doi.org/10.1145/3339825.3394933}{doi:\nolinkurl{10.1145/3339825.3394933}}


\bibitem[Nandakumar et~al\mbox{.}(2025)]%
        {nandakumar_mediaquictransport_2025}
\bibfield{author}{\bibinfo{person}{Suhas Nandakumar}, \bibinfo{person}{Victor Vasiliev}, \bibinfo{person}{Ian Swett}, {and} \bibinfo{person}{Alan Frindell}.} \bibinfo{year}{2025}\natexlab{}.
\newblock \bibinfo{booktitle}{\emph{Media over {{QUIC Transport}}}}.
\newblock \bibinfo{type}{Internet-{{Draft}}} draft-ietf-moq-transport-11. \bibinfo{institution}{Internet Engineering Task Force / Internet Engineering Task Force}.
\newblock
\urldef\tempurl%
\url{https://datatracker.ietf.org/doc/draft-ietf-moq-transport/11/}
\showURL{%
Retrieved 2025-09-08 from \tempurl}


\bibitem[Neugebauer(2022)]%
        {neugebauer_nagaremediaingest_2022}
\bibfield{author}{\bibinfo{person}{Matthias Neugebauer}.} \bibinfo{year}{2022}\natexlab{}.
\newblock \showarticletitle{Nagare {{Media Ingest}}: {{A Server}} for {{Live CMAF Ingest Workflows}}}. In \bibinfo{booktitle}{\emph{Proceedings of the 13th {{ACM Multimedia Systems Conference}}}}. \bibinfo{publisher}{ACM}, \bibinfo{address}{Athlone Ireland}, \bibinfo{pages}{210--215}.
\newblock
\showISBNx{978-1-4503-9283-9}
\href{https://doi.org/10.1145/3524273.3532888}{doi:\nolinkurl{10.1145/3524273.3532888}}


\bibitem[Pantos and May(2017)]%
        {pantos_httplivestreaming_2017}
\bibfield{author}{\bibinfo{person}{Roger Pantos} {and} \bibinfo{person}{William May}.} \bibinfo{year}{2017}\natexlab{}.
\newblock \bibinfo{title}{{{HTTP Live Streaming}}}.
\newblock \bibinfo{howpublished}{RFC 8216}.
\newblock
\href{https://doi.org/10.17487/RFC8216}{doi:\nolinkurl{10.17487/RFC8216}}


\bibitem[Parmar and Thornburgh(2012)]%
        {hparmar_adobesrealtime_2012}
\bibfield{author}{\bibinfo{person}{H. Parmar} {and} \bibinfo{person}{M. Thornburgh}.} \bibinfo{year}{2012}\natexlab{}.
\newblock \bibinfo{booktitle}{\emph{Adobe's {{Real Time Messaging Protocol}}}}.
\newblock \bibinfo{type}{{T}echnical {R}eport}. \bibinfo{institution}{Adobe}.
\newblock


\bibitem[Reschke(2015)]%
        {reschke_basichttpauthentication_2015}
\bibfield{author}{\bibinfo{person}{Julian Reschke}.} \bibinfo{year}{2015}\natexlab{}.
\newblock \bibinfo{title}{The '{{Basic}}' {{HTTP Authentication Scheme}}}.
\newblock \bibinfo{howpublished}{RFC 7617}.
\newblock
\href{https://doi.org/10.17487/RFC7617}{doi:\nolinkurl{10.17487/RFC7617}}


\bibitem[Sharabayko et~al\mbox{.}(2021)]%
        {sharabayko_srtprotocol_2021}
\bibfield{author}{\bibinfo{person}{Maria Sharabayko}, \bibinfo{person}{Maxim Sharabayko}, \bibinfo{person}{Jean Dube}, \bibinfo{person}{Joonwoong Kim}, {and} \bibinfo{person}{Jeongseok Kim}.} \bibinfo{year}{2021}\natexlab{}.
\newblock \bibinfo{booktitle}{\emph{The {{SRT Protocol}}}}.
\newblock \bibinfo{type}{Internet-{{Draft}}} draft-sharabayko-srt-01. \bibinfo{institution}{Internet Engineering Task Force}.
\newblock
\urldef\tempurl%
\url{https://datatracker.ietf.org/doc/html/draft-sharabayko-srt-01}
\showURL{%
Retrieved 2025-09-08 from \tempurl}


\bibitem[{Unified Streaming}(2022)]%
        {unifiedstreaming_unifiedstreamingfmp4ingest_2022}
\bibfield{author}{\bibinfo{person}{{Unified Streaming}}.} \bibinfo{year}{2022}\natexlab{}.
\newblock \bibinfo{title}{Unified {{Streaming}} fmp4ingest {{Tools DASH-IF Live Media Ingest Protocol}} - {{Interface}} 1 ({{CMAF}})}.
\newblock
\urldef\tempurl%
\url{https://github.com/unifiedstreaming/fmp4-ingest}
\showURL{%
Retrieved 2022-02-22 from \tempurl}


\bibitem[{Video Services Forum}(2020)]%
        {videoservicesforum_tr0612020reliable_2020}
\bibfield{author}{\bibinfo{person}{{Video Services Forum}}.} \bibinfo{year}{2020}\natexlab{}.
\newblock \bibinfo{booktitle}{\emph{{{TR-06-1}}:2020 {{Reliable Internet Stream Transport}} ({{RIST}}) {{Protocol Specification}} -- {{Simple Profile}}}}.
\newblock \bibinfo{type}{Standard}. \bibinfo{institution}{Video Services Forum}.
\newblock


\bibitem[{Video Services Forum}(2021a)]%
        {videoservicesforum_tr0622021reliable_2021}
\bibfield{author}{\bibinfo{person}{{Video Services Forum}}.} \bibinfo{year}{2021}\natexlab{a}.
\newblock \bibinfo{booktitle}{\emph{{{TR-06-2}}:2021 {{Reliable Internet Stream Transport}} ({{RIST}}) {{Protocol Specification}} -- {{Main Profile}}}}.
\newblock \bibinfo{type}{Standard}. \bibinfo{institution}{Video Services Forum}.
\newblock


\bibitem[{Video Services Forum}(2021b)]%
        {videoservicesforum_tr063reliableinternet_2021}
\bibfield{author}{\bibinfo{person}{{Video Services Forum}}.} \bibinfo{year}{2021}\natexlab{b}.
\newblock \bibinfo{booktitle}{\emph{{{TR-06-3 Reliable Internet Stream Transport}} ({{RIST}}) {{Protocol Specification}} -- {{Advanced Profile}}}}.
\newblock \bibinfo{type}{Standard}. \bibinfo{institution}{Video Services Forum}.
\newblock


\bibitem[{WHATWG}(2025)]%
        {whatwg_fetch_2025}
\bibfield{author}{\bibinfo{person}{{WHATWG}}.} \bibinfo{year}{2025}\natexlab{}.
\newblock \bibinfo{booktitle}{\emph{Fetch}}.
\newblock \bibinfo{type}{Living {{Standard}}}. \bibinfo{institution}{Web Hypertext Application Technology Working Group}.
\newblock
\urldef\tempurl%
\url{https://fetch.spec.whatwg.org/}
\showURL{%
Retrieved 2025-06-10 from \tempurl}


\end{thebibliography}
